%% file: main_clean.tex
\documentclass[aps,11pt,notitlepage,onecolumn,nofootinbib,superscriptaddress,longbibliography]{revtex4-2}

\input{pretex}
\usepackage{amsmath,mathrsfs,amsfonts,amssymb,array,graphicx,mathtools,multirow,bm,times,tcolorbox,relsize,booktabs}
\usepackage[utf8]{inputenc}
\usepackage[T1]{fontenc}
\usepackage{tikz} 
\usetikzlibrary{quantikz2}
\usepackage[ruled, vlined, linesnumbered]{algorithm2e}
\usepackage[vcentermath, enableskew]{youngtab}
\Yboxdim{7pt}

\definecolor{colortwo}{rgb}{0.4,0.77,0.17}
\definecolor{colorthree}{rgb}{0.01,0.51,0.93}

\nc{\EPPT}{{E_{\operatorname{PPT}}}}
\nc{\EPPTone}{{E_{\operatorname{PPT}}^{(1)}}}
\nc{\EK}{{E_{\kappa}}}
\nc{\Ext}{{{\operatorname{Ext}}}}
\nc{\sym}{{{\operatorname{sym}}}}
\allowdisplaybreaks

\begin{document}
\title{Quantifying Unextendibility via Virtual State Extension}
\author{Hongshun Yao}
\affiliation{Thrust of Artificial Intelligence, Information Hub,\\
The Hong Kong University of Science and Technology (Guangzhou), Guangdong 511453, China}
\author{Jingu Xie}
\affiliation{Thrust of Artificial Intelligence, Information Hub,\\
The Hong Kong University of Science and Technology (Guangzhou), Guangdong 511453, China}
\author{Xuanqiang Zhao}
\affiliation{QICI Quantum Information and Computation Initiative, School of Computing and Data Science, The University of Hong Kong, Pokfulam Road, Hong Kong}
\author{Chengkai Zhu}
\affiliation{Thrust of Artificial Intelligence, Information Hub,\\
The Hong Kong University of Science and Technology (Guangzhou), Guangdong 511453, China}
\author{Ranyiliu Chen}
\email{chenranyiliu@quantumsc.cn}
\affiliation{Quantum Science Center of Guangdong-Hong Kong-Macao Greater Bay Area, Shenzhen 518045, China}
\author{Xin Wang}
\email{felixxinwang@hkust-gz.edu.cn}
\affiliation{Thrust of Artificial Intelligence, Information Hub,\\
The Hong Kong University of Science and Technology (Guangzhou), Guangdong 511453, China}

\date{\today}

\begin{abstract}
Monogamy of entanglement, which limits how entanglement can be shared among multiple parties, is a fundamental feature underpinning the privacy of quantum communication. In this work, we introduce a novel operational framework to quantify the unshareability or unextendibility of entanglement via a virtual state-extension task. The virtual extension cost is defined as the minimum simulation cost of a randomized protocol that reproduces the marginals of a $k$-extension. For the important family of isotropic states, we derive an exact closed-form expression for this cost. Our central result establishes a tight connection: the virtual extension cost of a maximally entangled state equals the optimal simulation cost of universal virtual quantum broadcasting. Using the algebra of partially transposed permutation matrices, we obtain an analytical formula and construct an explicit quantum circuit for the optimal broadcasting protocol, thereby resolving an open question in quantum broadcasting. We further relate the virtual extension cost to the absolute robustness of unextendibility, providing it with a clear operational meaning, and show that the virtual extension cost is an entanglement measure that bounds distillable entanglement and connects to logarithmic negativity.
\end{abstract}

\maketitle

\tableofcontents
\section{Introduction}

Entanglement is one of the defining features of quantum mechanics and an indispensable resource for quantum information processing \cite{Horodecki2009}. Quantifying entanglement remains a central challenge in quantum information theory, particularly when moving beyond the idealized, asymptotic setting. In recent years, attention has shifted toward operationally meaningful and computationally tractable entanglement measures that capture the performance limits of quantum technologies in realistic, finite-copy regimes. A promising direction in this line of work has been to characterize entanglement via its \emph{(non-)extendibility} properties, leading to the development of the resource theory of unextendibility (see, e.g.,~\cite{kaur2019extendibility,kaur2021resource,wang2024quantifying}), for which we treat $k$-extendible states~\cite{Werner1989,Doherty_2002,Doherty_2004} as free states.

Specifically, unextendibility leverages the structure of symmetric extensions to capture the ``monogamy of quantum entanglement'' \cite{Coffman2000,Terhal2004,Koashi2004,Yang2006,Osborne2006,Adesso2007b,Lancien2016} in a quantitative way. For example, any three parties cannot share the maximally entangled states between each two of them at the same time. However, inspired by the recent development of \textit{virtual} protocols \cite{Temme2017,van_den_Berg_2023,Wang2020,yuan2024virtual,parzygnat2024virtual}, one may ask that can we go beyond the entanglement monogamy in a virtual way?
In this work, we introduce the task of \emph{virtual state extension} to answer this affirmatively. Concretely, given a potentially unextendible bipartite state $\rho_{AB}$, we ask how hard it is to virtually prepare a $k$-extension of it (a hermitian operator on $A B_1 B_2 \cdots B_k$ whose marginal on any $A B_i$ is equal to $\rho_{AB}$). We then define the cost of virtual extension (or the implementibility \cite{jiang2021physical} of the preparation HPTP map)
as a measure of unextendibility. By construction, this measure vanishes exactly for $k$-extendible states.

We provide closed-form expressions for the virtual extension cost for the family of isotropic states. This requires new analytic tools leveraging the walled Brauer algebra and mixed Schur–Weyl duality and could be of independent algebraic interest. In particular, the special case of the maximally entangled state proves the most instructive, which connects directly to the universal virtual broadcasting protocol of \cite{yao2024optimal}. Indeed, if one could (virtually) broadcast the maximally entangled state and then perform teleportation, one would effectively obtain a universal broadcast of arbitrary states. We formalize this connection and, moreover, construct an explicit circuit implementation of the optimal universal virtual‐broadcasting protocol, which was left open in \cite{yao2024optimal}. This construction thus constitutes a further contribution of our work.

We next adopt a resource-theoretic perspective to analyze the virtual extension cost. In this framework, we show that it is equivalent to the absolute robustness of extendibility, thus endowing the latter with an operational meaning. (We note that in \cite{kaur2021resource} the authors studied the \emph{global} robustness of extendibility, which is distinct from our notion.) From that equivalence, properties such as faithfulness and monotonicity follow naturally. We further compare our measure with known entanglement quantifiers and show that it has an analytic form for bipartite pure states that equals the (conventional) logarithmic negativity when $k\to\infty$. Also, it differs from the global robustness of extendibility \cite{kaur2021resource} on some explicit state. Finally,
it yields nontrivial upper bounds on one-shot 1‑LOCC exact entanglement distillation.

Taken together, our contributions offer a fresh, operationally grounded lens on unextendibility and its interplay with entanglement theory by linking the explicit (although nonphysical) state extension task to the resource-theoretic constructions. Our approach opens a pathway to understanding the interplay between entanglement, symmetries, and representation theory in constrained quantum information tasks.
\section{Preliminaries}
\subsection{Notations}
Throughout this paper,  
finite-dimensional Hilbert spaces where quantum systems $A$, $B,\dots$ underlie are denoted by $\mathcal{H}_A, \mathcal{H}_B, \dots$, 
with quantum states represented as density operators $\rho, \sigma \in \mathscr{D}(\mathcal{H})$, where  $\mathscr{D}(\mathcal{H})$ represents the set of all density operators on $\mathcal{H}$.  The trace norm is defined as $\|\rho\|_1 :=\tr\sqrt{\rho^\dagger\rho}$, where the symbol $\dagger$ in the upper right corner denotes the conjugate transpose.
For a bipartite system, the normalized maximally entangled state of local dimension $d$ is written as 
$\Phi_{AB}^d = \tfrac{1}{d} \sum_{i,j} |ii\rangle \langle jj| $ and the unnormalized one as $\Psi_{AB}^d = \sum_{i,j} |ii\rangle \langle jj| $. In the multi-system setting, we use the shorthand $B^k$ to denote the composite system $B_1B_2\cdots B_k$. For a multipartite state, the notation $\mathrm{Tr}_{\backslash AB_j}$ stands for the partial trace over all subsystems except $A$ and $B_j$.
The transpose of an operator is indicated by $(\cdot)^T$, while the partial transpose 
on subsystem $B$ is denoted by $T_B$. For multipartite systems, $T_m$ indicates the transposition on the last $m$ subsystems. The identity channel is denoted by $\mathrm{id}$. 
For a quantum channel $\mathcal{N}_{A\to B}:\mathscr{D}(\cH_A)\to\mathscr{D}(\cH_B)$, its Choi operator is written as $J^{\mathcal{N}}_{A\to B}:=(\mathrm{id}\otimes\mathcal{N})(\Psi^{\dim A}_{AB})$.

In the table below, we summarize the notations used throughout the paper:
\setlength\extrarowheight{2.2pt}
\begin{table}[H]
\centering
\begin{tabular}{l|l}
\toprule[2pt]
Symbol & Definition\\
\hline
$A,A_j,A^\prime,B,B_j,B^\prime,\cdots$ & Different quantum systems\\
$A^n,B^n,\cdots$ & Abbreviation for quantum systems $A_1\cdots A_n$ and $B_1\cdots B_n,\cdots$\\
{$\cH_A,{\cH_{B}},\cH_{A^n}\cdots$} & {Hilbert space of quantum system $A,B,A^n\cdots$} \\
{$\mathscr{D}(\cH_A)$} & {Set of density matrices on $\cH_A$} \\
{$\rho, \sigma$} & Quantum states in $\mathscr{D}(\cH_A)$\\
$\Phi_{AB}^d$ & $d\times d$ maximally entangled state on systems $A$ and $B$\\
$\Psi_{AB}^d$ & Choi operator of identity map\\
$\NN_+$ & Set of natural number\\
$\id$ & {Identity map on the space $\mathscr{D}(\cH_A)$}\\
$I$ & Identity operator\\
$J^{\cN}_{A\to B}$ & Choi representation of the map $\cN_{A\to B}$\\
$\cN_1,\cN_2,\cdots$ & Completely Positive Trace-Preserving (CPTP) maps\\
$\Lambda,\Gamma,\cdots$ & Hermitian-Preserving Trace-Preserving (HPTP) maps\\
$(\cdot)^T$ & Transpose of an operator\\
$\mathbb{Y}_n^d$ & The set of Young Diagrams\\
$\lambda,\mu,\nu, \alpha,\beta$ & Partions/Young Diagrams\\
$\textbf{S}_n$ & Symmetric group on $n$ elements\\
$\textbf{SU}(d)$ & Unitary group\\
$P^{\lambda}$ & Young projector\\
$L(\tau)$ &  The length of the cycle containing position 1 for a fixed \(\tau \in S_n\)\\
$c(\tau)$ &  The  number of cycles for a fixed \(\tau \in S_n\)\\
\bottomrule[2pt]  
\end{tabular}
\caption{\small Overview of notations. 
}
\label{tab: state version}
\end{table}

\subsection{Elements for group representation theory}
This section presents the mathematical foundations for deriving the main result, including the partitions of natural numbers, the group representation of the symmetry group $\textbf{S}_n$, and the unitary group $\textbf{SU}(d)$. Specifically, we focus on the duality relationship between these two groups, i.e., the Schur-Weyl duality and Mixed Schur-Weyl duality~\cite{bacon2006efficient,nguyen2023mixed}. A crucial observation is that any operator commuting with tensor representations of the unitary group belong to the symmetric group algebra, whereas those commuting with tensor representations of mixed unitary groups belong to the Walled Brauer algebra~\cite{benkart1994tensor}. These symmetries directly facilitate the simplification of the entanglement-theoretic problems we aim to analyze.
\subsubsection{Young diagram}
For a given natural number $n$, we define its \emph{partition} $\lambda$ as a $n$-tuple of positive numbers $\lambda:=(\lambda_1,\cdots,\lambda_d)$, such that
\begin{equation}
    \begin{aligned}
        \lambda_1\geq\cdots\geq\lambda_d\geq 0\,\text{ and}\,\sum_{i=1}^d\lambda_i=n.
    \end{aligned}
\end{equation}
Let $\mathbb{Y}_n^d:=\{\lambda|\lambda_1\geq\cdots\geq\lambda_d\geq 0\,\text{and}\,\sum_{i=1}^d\lambda_i=n\}$ denote the partition set of $n$. Every partition $\lambda\in \mathbb{Y}_n^d$ can be visualized as a Young Diagram in which there are up to $d$ rows with $\lambda_i$ boxes in row $i$. For example, for partitions $\lambda=(2,1)$ and $\mu=(3,2,2)$, we have the Young Diagrams as follows:
\begin{equation}
    \begin{aligned}
        \lambda=\yng(2,1),\quad \mu=\yng(3,2,2).
    \end{aligned}
\end{equation}
Then, we further define a Standard Young tableau $\text{T}$ of shape $\lambda$ to be a way of filling the $n$ boxes of $\lambda$ with the integers $1,\cdots,n$, using each number once and so that integers increase from left to right and from top to bottom. We define the set of Standard Young tableau as $\text{SYT}(\lambda,d)$. There is another way to fill the boxes of a Young diagram. For a shape $\lambda\in \mathbb{Y}_n^d$, if the boxes are filled with numbers $1,\cdots,d$ so that the integers are increasing from top to bottom in each column and non-decreasing from left to right in each row, such a Young tableau is called a Semistandard Young tableau. Similarly, we denote the corresponding tableau set as $\text{SSYT}(\lambda, d)$.

In this paper, we write $\lambda+\yng(1)$ as the Young Diagram set in which each element can be obtained from a Young Diagram $\alpha\in\mathbb{Y}_n^d$ by adding a one box. For example, suppose $\lambda=(2,1)\in\mathbb{Y}_3^d$, we have
\begin{equation}
    \begin{aligned}
        \lambda+\yng(1):=\left\{\beta:=\yng(3,1),\,\mu:=\yng(2,2),\,\nu:=\yng(2,1,1)\right\}.
    \end{aligned}
\end{equation}
Note that if the system dimension $d\leq 2$, the last Young Diagram $\nu$ in the set $\lambda+\yng(1)$ must be removed.
\subsubsection{Schur-Weyl duality}
The Schur-Weyl duality characterizes the dual relationship between representations of the symmetric group and the unitary group. Expressed in matrix terms, the representation matrices of these two groups can be simultaneously block-diagonalized under the same basis, known as the Schur basis, with this transformation referred to as the Schur transform.

Specifically, we have the following natural representation of the symmetric group on the space $(\mathbb{C}^d)^{\otimes n}$:
\begin{equation}
    \begin{aligned}
        \textbf{P}(\sigma)\ket{i_1}\otimes\cdots\otimes\ket{i_n}=\ket{i_{\sigma^{-1}(1)}}\otimes\cdots\otimes\ket{i_{\sigma^{-1}(n)}},
    \end{aligned}
\end{equation}
where $\textbf{P}(\sigma):=V(\sigma)$ is the permutation operator for $\sigma\in \textbf{S}_n$ and $\sigma(i)$ is the label describing the action of $\sigma$ on label $i$. For example, for $\pi=(123)$, we have $\textbf{P}(\sigma)\ket{i_1,i_2,i_3}=\ket{i_3,i_1,i_2}$. Denote $(\textbf{P}(\cdot),(\mathbb{C}^d)^{\otimes n})$ as representation of symmetric group $\textbf{S}_n$. This representation naturally extends to the representation of the group algebra $\mathbb{C}\textbf{S}_n$. Similarly, the $n$-fold tensor representation of the unitary group $\textbf{SU}(d)$ is given by
\begin{equation}
    \begin{aligned}
        \textbf{Q}(U)\ket{i_1}\otimes\cdots\otimes\ket{i_n}=U\ket{i_1}\otimes\cdots\otimes U\ket{i_n},
    \end{aligned}
\end{equation}
where $\textbf{Q}(U):=U^{\otimes n}$ for all $U\in\textbf{SU}(d)$. Denote $(\textbf{Q}(\cdot),(\mathbb{C}^d)^{\otimes n})$ as representation of unitary group $\textbf{SU}(d)$. It is straightforward to see that the representations commute, i.e., $\textbf{P}(\sigma)\textbf{Q}(U)=\textbf{Q}(U)\textbf{P}(\sigma)$. Consequently, both the Hilbert space $(\mathbb{C}^d)^{\otimes n}$, and operators $\textbf{P}(\sigma)$ and $\textbf{Q}(U)$ can be decomposed as
\begin{equation}
    \begin{aligned}
         \textbf{Q}(U)\textbf{P}(\sigma)\overset{\textbf{SU}(d)\times \textbf{S}_n}{\cong}\bigoplus_{\lambda\in \mathbb{Y}_n^d} \textbf{q}_{\lambda}(U)\otimes \textbf{p}_{\lambda}(\sigma),\quad (\mathbb{C}^d)^{\otimes n}\overset{\textbf{SU}(d)\times \textbf{S}_n}{\cong}\bigoplus_{\lambda\in \mathbb{Y}_n^d} \mathbb{Q}_{\lambda}\otimes \mathbb{P}_\lambda,
    \end{aligned}
\end{equation}
where $(\textbf{p}_{\lambda},\mathbb{P}_{\lambda})$ and $(\textbf{q}_\lambda,\mathbb{Q}_\lambda)$ denote the irreducible representation labeled by the Young Diagram $\lambda$. Denote $m_\lambda:=\text{dim}(\mathbb{P}_\lambda)$ and $d_\lambda:=\text{dim}(\mathbb{Q}_\lambda)$. There are some facts that dimensions $d_\lambda$ and $m_\lambda$ are equal to the number of semistandard Young tableaux, i.e., $d_\lambda=|\text{SSYT}(\lambda,d)|$ and standard Young tableaux, i.e., $m_\lambda=|\text{SYT}(\lambda,d)|$, respectively.

By Schur's lemma, any operator commuting with $U^{\otimes n}$ for all $U\in\textbf{SU}(d)$ can be spanned by matrix basis $E^\lambda_{ij}$ defined by
\begin{equation}
    \begin{aligned}
        E^\lambda_{ij}:=\mathbb{I}_{\mathbb{Q}_\lambda}\otimes\ketbra{\lambda,i}{\lambda,j}_{\mathbb{P}_\lambda},\quad \lambda\in\mathbb{Y}_n^d,\,i,j\in\{1,\cdots,m_\lambda\},
    \end{aligned}
\end{equation}
where $\{\ket{\lambda,i}_{\mathbb{P}_\lambda}\}$ is the Young-Yamanouchi basis. There are some rules related to this basis:
\begin{equation}
    \begin{aligned}
        E^\lambda_{ij}E^{\mu}_{kl}=\delta_{\lambda\mu}\delta_{jk}E^\lambda_{il},\quad \tr[E^\lambda_{ij}]=d_\lambda\delta_{ij}.
    \end{aligned}
\end{equation}
The Young projector is further defined based on the above basis with following properties:
\begin{equation}
    \begin{aligned}
        P^\lambda:=\sum_{i=1}^{m_\lambda}E^\lambda_{ii},\quad P^\lambda P^\mu=\delta_{\lambda\mu}P^\lambda,\quad \tr[P^\lambda]=d_\lambda m_\lambda.
    \end{aligned}
\end{equation}
\subsubsection{Partially transposed permutations matrix algebra}
For any unitary $U\in\textbf{SU}(d)$, the operators commuting with $U^{\otimes n}$ are elements of the group algebra $\mathbb{C}\textbf{S}_n$, while those commuting with $U^{\otimes n}\otimes \overline{U}^{\otimes m}$ belong to the walled Brauer algebra $\cB_{n,m}^d$. Within the framework of Schur-Weyl duality, our focus lies on the representations of the group algebra $\mathbb{C}\textbf{S}_n$, that is, the algebra composed of permutation operators. In this section, we delve into the Partially transposed permutations matrix algebra $\cA_{n,m}^d$, which is the representation of the walled Brauer algebra. Then, the dual properties and corresponding transformations are referred to as mixed Schur-Weyl duality and mixed Schur transform, respectively.

Let $n,m,d\in\mathbb{\NN}^+$. The walled Brauer algebra $\cB_{n,m}^d$ consists of complex linear combinations of \textit{diagrams}, where each diagram has two rows of $n + m$ nodes each, with a vertical “wall” between the first $n$ and the last $m$ nodes. The nodes are connected into pairs, subject to the restriction: if both nodes are in the same row, they must be on different sides of the wall, while if they are in different rows, they must be on the same side of the wall. See the following example.
\begin{figure}[H]
    \centering
    \includegraphics[width=0.27\linewidth]{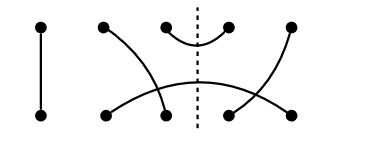}
    \caption{An element in the walled Brauer algebra $\cB_{3,2}^d$~\cite{nguyen2023mixed}.}
'    \label{fig:walled_brauer_algebra_3_2}
\end{figure}
The walled Brauer algebra $\cB_{n,m}^d$ is a partial transpose (on the last $m$ systems) of the symmetric group algebra $\mathbb{C}\mathbf{S}_{n+m}$. Specifically, we have the following connection between these two algebras:

\begin{figure}[H]
    \centering
    \includegraphics[width=0.55\linewidth]{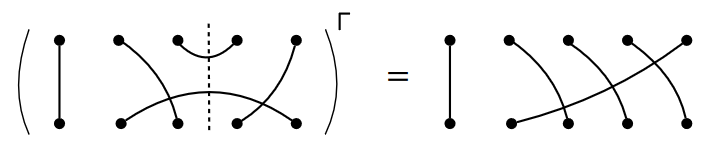}
    \caption{Partial transpose $\sigma^\Gamma$ of a diagram $\sigma$ in $\cB_{3,2}^d$ corresponds to exchanging the last $2$ nodes~\cite{nguyen2023mixed}.}
    \label{fig:walled_brauer_algebra_symmetrc_group}
\end{figure}
Then, the representation of the walled Brauer algebra is given by 
\begin{equation}
    \begin{aligned}
        \cA_{n,m}^d:=\text{span}_{\mathbb{C}}\{\textbf{P}_{n,m}(\sigma)|\sigma\in\cB_{n,m}^d\}=\text{span}_{\mathbb{C}}\{\textbf{P}(\sigma)^{T_{m}}|\sigma\in\textbf{S}_{n+m}\},
    \end{aligned}
\end{equation}
where $T_m$ denotes the partial trace on the last $m$ systems, and $\textbf{P}_{n,m}$ denotes the algebra representation, which is called \textit{Partially transposed permutation matrix algebra}.

In this work, we focus on the specific case, $m=1$. Note that the algebra $\cA_{n,1}^d$ can be represented as a direct sum of two ideals~\cite{mozrzymas2014structure,studzinski2017port}, i.e.
\begin{equation}\label{eq:ideal_decomposition}
    \begin{aligned}
        \cA_{n,1}^d=\cM\oplus\cS=F\cA_{n,1}^dF+(id_{\cA}-F)\cA_{n,1}^d(id_{\cA}-F),
    \end{aligned}
\end{equation}
where $F=\sum_{\alpha\in\mathbb{Y}_{n-1}^d}\sum_{\mu=\alpha+\yng(1)}F_{\mu}(\alpha)$ is the identity on the ideal $\cM$ and $id_\cA$ is the identity operator on the whole space. The operators $F_\mu(\alpha)$ are projectors on the irreps of $\cA_{n,1}^d$ contained in the ideal $\cM$ and they satisfy the following rules
\begin{equation}\label{eq:projectors_ideal_C}
    \begin{aligned}
        F_{\mu}(\alpha)F_{\nu}(\beta)=\delta_{\mu\nu}\delta_{\alpha\beta}F_\mu(\alpha),\quad\tr(F_\mu(\alpha))=d_\alpha m_\mu.
    \end{aligned}
\end{equation}
The explicit form of the projectors $F_\mu(\alpha)$ in natural representation is given by
\begin{equation}
    \begin{aligned}
        F_\mu(\alpha)=\frac{d_\alpha m_\mu}{nd_\mu m_\alpha} P_\mu\sum_{k=1}^{n}S_{k,n}P_\alpha S_{n,n+1}^{T_{n+1}} S_{k,n} P_\mu,
    \end{aligned}
\end{equation}
where $S$ is the swap operator, $P_\alpha$ and $P_\mu$ are Young projectors onto irreducible spaces labelled by the Young diagrams $\alpha\in\mathbb{Y}_{n-1}^d$ and $\mu\in\mathbb{Y}_{n}^d$, respectively. The identity operators are omitted.
\subsection{Quantum virtual broadcasting}
The no-broadcasting theorem states that no physical operation, mathematically referred to as a completely positive and trace-preserving (CPTP) map, can broadcast an unknown quantum state to two parts such that each part attains the original state~\cite{barnum1996noncommuting,kalev2008no}.
Specifically, when $\mathcal{H}_A\cong\mathcal{H}_{B_1}\cong\mathcal{H}_{B_2}$ there is no quantum channel $\mathcal{B}: \mathscr{D}(\cH_A) \to \mathscr{D}(\cH_{B_1}\ox\cH_{B_2})$ such that
\begin{equation}\label{eq:broadcasting}
    \tr_{B_1}[\mathcal{B}(\rho)] = \tr_{B_2}[\mathcal{B}(\rho)] = \rho \quad \forall \rho \in \mathscr{D}(\cH_{A}).
\end{equation}
Nonetheless, Eq.~\eqref{eq:broadcasting} becomes feasible when the map $\cB$ is relaxed to Hermitian-preserving and trace-preserving (HPTP) maps, and the notion of \textit{Quantum Virtual Broadcasting}~\cite{parzygnat2024virtual,yao2024optimal,xiao2025no,xiao2025nogf} was introduced.
\textit{Universal Virtual Broadcasting} refers to HPTP maps that broadcast any state~\cite{yao2024optimal}, whose efficacy are gauged by their simulation cost \cite{jiang2021physical}. The universal virtual broadcasting protocol that minimizes this simulation cost is deemed optimal.

\begin{definition}[Simulation cost of an HPTP map~\cite{jiang2021physical}]\label{def:cost_of_hptp}
The simulation cost (or physical implementability) of an HPTP map $\Gamma$ is defined as
\begin{equation}
\begin{aligned}
    \nu(\Gamma):=\log\min\Big\{p_1+p_2\big| \, \Gamma = p_1\mathcal{N}_1-p_2\mathcal{N}_2,~p_1,p_2\geq 0, \cN_1,\cN_2\in\text{CPTP} \Big\}.
\end{aligned}
\end{equation}
\end{definition}

By Hoeffding's inequality, for any observable $O$ and state $\rho$, at least $\cO((\frac{2^{\nu(\Gamma)}}{\delta})^2\ln{\frac{2}{\epsilon}})$ samples are required to estimate $\tr[O\Gamma(\rho)]$ for an error less than $\delta$, with probability $1-\epsilon$. In this sense, $\nu(\cdot)$ characterizes the resource cost required to simulate non-physical operations using physical operations.

\begin{definition}[\cite{yao2024optimal}]\label{def:optimal simulation cost}
    For quantum systems $A$ and $B^k$, the optimal simulation cost of universal $k$-broadcasting protocols from $A$ to $B^k$ is defined by
    \begin{align}
        \gamma_k:=\min\{\nu(\Gamma_{A\to B^k}) ~\vert~ \Gamma_{A\to B^k}\in\mathcal{T}_k\},
    \end{align}
    where $\mathcal{T}_k$ denotes the set of all universal virtual $k$-broadcasting protocols $A\to B^k$.
\end{definition}

Generally, we also incorporate probabilities $p_1$ and $p_2$ into $\cN_1$ and $\cN_2$, respectively, and uniformly treat them as CPTN maps for analysis. Then, the optimal simulation cost of all universal virtual $k$-broadcasting protocols can be characterized as the following SDP~\cite{yao2024optimal}:
\begin{equation}\label{eq:primal_without_rho}
\begin{aligned}
    2^{\gamma_k} =  \min\;&p_1+p_2\\
    {\rm s.t.}\; &\tr_{\backslash AB_j}[J^{\cN_{1}}_{A\to B^k}-J^{\cN_{2}}_{A\to B^k}]=\Psi_{AB_j}^d,\,j=1,\cdots,k\\
    &\tr_{B^k}[J^{\cN_{1}}_{B\to B^k}]=p_1I_{A},\\
    &\tr_{B^k}[J^{\cN_{2}}_{B\to B^k}]=p_2I_{A},\\
    &J^{\cN_{1}}_{B\to B^k}\geq 0, J^{\cN_{2}}_{B\to B^k}\geq 0,
\end{aligned}
\end{equation}
where $\Psi_{AB_j}^d$ is the unnormalized $d\otimes d$ maximally entangled state on system $AB_j$.
\subsection{Resource theory of unextendibility}
Unextendibility or unshareability of entanglement reflects the idea that the more entangled a bipartite state is, the less entangled its individual systems can be with a third party. Recall the definition of $k$-extendible states \cite{vedral1998entanglement,doherty2002distinguishing,Doherty_2004}: for any $k\in\NN_+$, a bipartite state $\rho_{AB}$ is \emph{$k$-extendible} if there exists a state $\sigma_{AB_1\cdots B_k}$ that satisfies
\begin{align}
    \rho_{AB}&=\tr_{\setminus AB_j}[\sigma_{AB_1\cdots B_k}]
\end{align}
for all $j\leq k$. The set of $k$-extendible states is denoted by ${\rm EXT}_k$.

Unextendibility provides a handy way to detect entanglement: if $\rho$ is separable, then $\rho$ is $k$-extendible for all $k$; if $\rho$ is entangled, then it is not $k$-extendible for some $k$, hence not $k'$-extendible for all $k'>k$.
Fitting unextendibility into the resource theory framework \cite{gour2024resourcesquantumworld}, one can define the \emph{robustness of unextendibility} as the following:

\begin{definition}[Robustness of unextendibility]
    For a given quantum state $\rho$, the \emph{absolute robustness of $k$-unextendibility} is defined as 
    \begin{equation}
        \begin{aligned}
            \cR_{\rm{absolute}}^{{\rm EXT}_k}(\rho):=\inf\left\{s~\middle\vert~\sigma\in{\rm EXT}_k,\frac{\rho+s\sigma}{1+s}\in{\rm EXT}_k\right\},
        \end{aligned}
    \end{equation}
and the \emph{global robustness $k$-unextendibility} is defined as 
    \begin{equation}
        \begin{aligned}
            \cR_{\rm{global}}^{{\rm EXT}_k}(\rho):=\inf\left\{s~\middle\vert~\sigma\in\mathscr{D}(\cH),\frac{\rho+s\sigma}{1+s}\in{\rm EXT}_k\right\}.
        \end{aligned}
    \end{equation}
    \label{def:rob}
\end{definition}

It holds that $\cR_{\rm{global}}^{{\rm EXT}_k}(\rho)=\inf_{\sigma\in{\rm EXT}_k}D_{\max}(\rho\|\sigma)-1$ \cite[Lemma 10.1]{gour2024resourcesquantumworld}, where the $k$-unextendible max-relative entropy $\inf_{\sigma\in{\rm EXT}_k}D_{\max}(\rho\|\sigma)$ was introduced in \cite{kaur2019extendibility,kaur2021resource}. In contrast, the quantity $\cR_{\rm{absolute}}^{{\rm EXT}_k}(\rho)$ receives comparatively little attention.
\section{Main results}
\subsection{Virtual state $k$-extension}

\begin{figure}[H]
    \centering
    \includegraphics[width=0.7\linewidth]{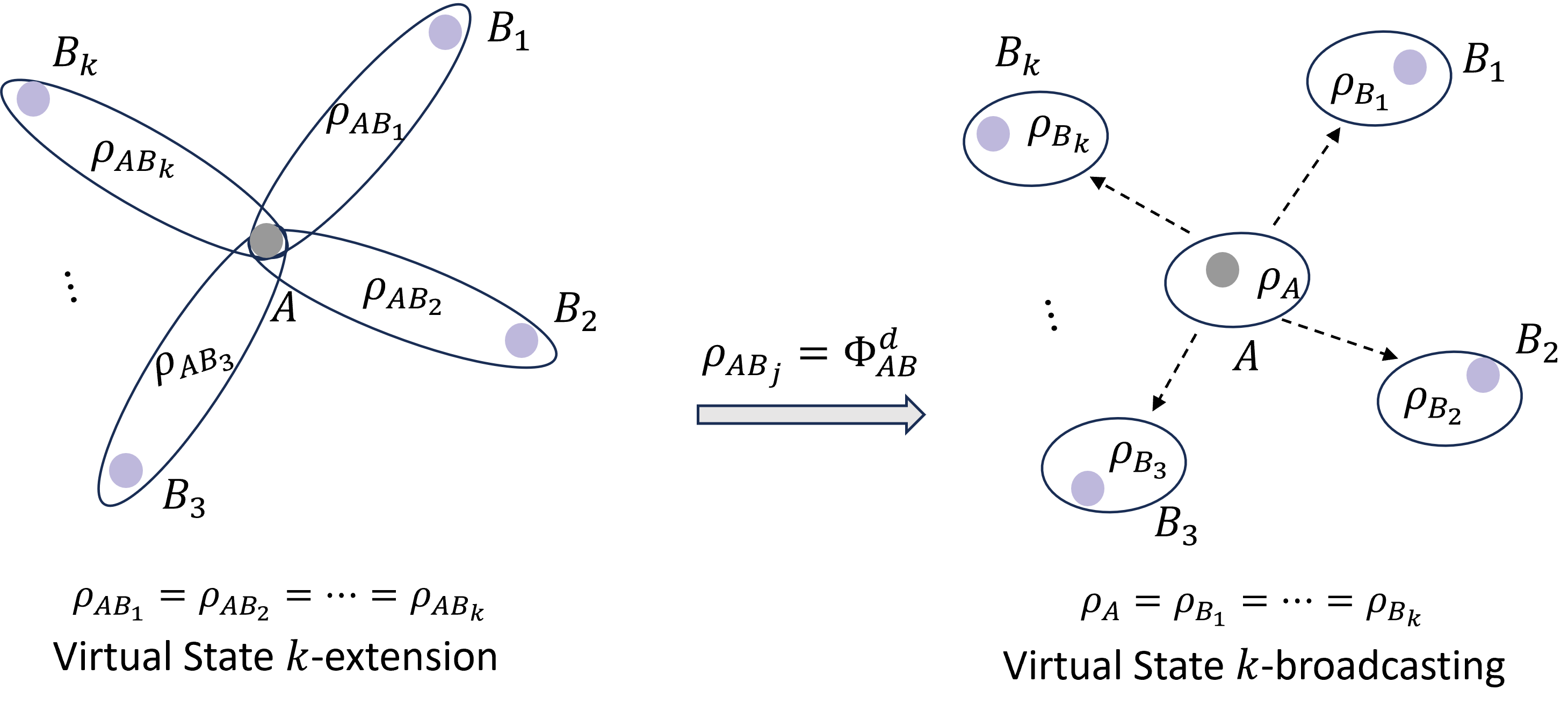}
    \caption{Schematic diagram of virtual state $k$-extension (left) and virtual state $k$-broadcasting (right). The central grey dot represents the central hub, while the surrounding purple dots denote information stations. In the left diagram, the hub and each station share the same state $\rho_{AB_j}$ statistically. When this state is maximally entangled, i.e. $\Phi_{AB}^d$, the protocol effectively becomes universal virtual broadcasting (right), where the hub broadcasts a state to the stations.}
    \label{fig:ext_broadcasting}
\end{figure}

This section introduces the core concept of our work: the virtual state $k$-extension. We begin by defining the virtual extension cost $\eta_k(\rho_{AB})$ of a bipartite state $\rho_{AB}$ as the minimum simulation cost of Hermitian-preserving and trace-preserving (HPTP) maps that generate a $k$-extension of $\rho_{AB}$. 
To gain further insight, we then solve the case of the maximally entangled state $\Phi^d$. Remarkably, the virtual extension problem for $\Phi^d$ turns out to be equivalent to the task of universal virtual broadcasting for arbitrary quantum states: intuitively, extending a maximally entangled state enables broadcasting of an arbitrary state by teleportation. 
Proposition~\ref{prop:vse_to_vb} formalizes this equivalence and highlights the conceptual bridge it builds between entanglement theory and quantum communication theory (see Fig.~\ref{fig:ext_broadcasting}).

\begin{shaded}
\begin{definition}[Virtual state extension]
For a bipartite state $\rho_{AB}\in\mathscr{D}(\cH_A\otimes \cH_B)$ and an integer $k \in \mathbb{N}_{+}$, an HPTP map $\Gamma
_{AB\to AB^k}$ is called a virtual state $k$-extension protocol for $\rho_{AB}$ if
\begin{equation}
    \rho_{AB}=\mathrm{Tr}_{\backslash AB_j}[\Gamma_{AB\to AB^k}(\rho_{AB})], \quad \forall j \in \{1, \dots, k\},
\end{equation}
where $\tr_{\backslash AB_j}$ denotes taking partial trace on subsystems excluding $AB_j$, and $B^k$ is the abbreviation of the subsystems $B_1B_2\cdots B_k$.
\end{definition}
\end{shaded}

\begin{shaded}
\begin{definition}[Virtual extension cost]\label{def:cost_of_vse}
For a given bipartite state $\rho_{AB}\in\mathscr{D}(\cH_A\otimes \cH_B)$ and an integer $k \in \mathbb{N}_{+}$, the optimal simulation cost of all virtual state $k$-extension protocols for $\rho_{AB}$ is defined as
\begin{equation}
    \begin{aligned}
        \eta_k(\rho_{AB}):=\min\{\nu(\Gamma_{AB\to AB^k}):\Gamma_{AB\to AB^k}\in \cE_{k,\rho_{AB}}\},
    \end{aligned}
\end{equation}
where $\cE_{k,\rho_{AB}}$ denotes the set of all virtual state $k$-extension protocols for $\rho_{AB}$. $\nu(\cdot)$ can be found in Def.~\eqref{def:cost_of_hptp}. We refer to the optimal simulation cost as the virtual extension cost.
\end{definition}
\end{shaded}

\begin{tcolorbox}
\begin{proposition}\label{prop:vse_to_vb} Virtual state $k$-extension {cost} for maximally entangled state $\Phi^{d}$ is equivalent to the {optimal simulation cost of} universal virtual $k$-broadcasting (Def.~\eqref{def:optimal simulation cost}), i.e.,
\begin{equation}
    \eta_{k}(\Phi^{d})=\gamma_{k}.
\end{equation}
\end{proposition}
\end{tcolorbox}
\begin{proof}
    Note that instead of optimizing over HPTP maps in the definition, one can equivalent express $\eta_k(\rho_{AB})$ as optimization over Hermitian operators $M_{AB^k}$:
    \begin{equation}
        \begin{aligned}\label{eq:cost_as_state_opt}
            2^{\eta_k(\rho_{AB})} = \min\{\Vert M_{AB^k} \Vert_1 : \rho_{AB} = \mathrm{Tr}_{\backslash AB_j}[M_{AB^k}] \quad \forall j \in \{1, \dots, k\}\}.
        \end{aligned}
    \end{equation}
    By the SDP of the trace norm, it is straightforward to obtain the following optimization problem:
    \begin{equation}
        \begin{aligned}
            2^{\eta_k(\Phi^d)}=  \min\;&\tr[Q]+\tr[S]\\
    {\rm s.t.}\; &\tr_{\backslash AB_j}[Q-S]=\Phi^d,\,j=1,\cdots,k\\
    &Q\geq 0, S\geq 0.
        \end{aligned}
    \end{equation}
    Since the operators $\tr_{\backslash AB_j}[Q]$ and $\tr_{\backslash AB_j}[S]$ for any $j=1,\cdots,k$ possess the same symmetry as the maximally entangled state $\Phi^d$, i.e., commuting with $U\otimes \overline{U}$ for all $U\in \textbf{SU}(d)$. It means that they can be expressed as linear combinations of identity operators and maximally entangled states. As a result, one can assume $\tr_{\backslash A}[Q]=\frac{p_1}{d} I$ and $\tr_{\backslash A}[S]=\frac{p_2}{d} I$, and introduce new optimization variables $p_1$ and $p_2$. By replacing $Q$ and $S$ with $J^{\cN_1}/d$ and $J^{\cN_2}/d$, respectively, we obtain the SDP for universal virtual broadcasting in Eq.~\eqref{eq:primal_without_rho}, which completes this proof.
\end{proof}
\subsubsection{Optimal $k$-broadcasting protocol} 
The equivalence $\eta_k(\Phi^d) = \gamma_k$ implies that solving for the fundamental cost of virtually extending a maximally entangled state is the same as finding the optimal protocol for broadcasting an arbitrary quantum state. As shown in Eq.~\eqref{eq:primal_without_rho}, this cost can be calculated by a semidefinite program (SDP). While SDPs can be solved numerically, an analytical, closed-form solution provides deeper insight.

The key lies in exploiting the inherent symmetries of the extension task. The protocol must be "universal," meaning it is covariant with respect to unitary transformations of the input state. Furthermore, the output must consist of identical marginals, implying a permutation symmetry among the output systems. These symmetries—unitary covariance and permutation invariance—are precisely captured by the algebraic framework of Schur-Weyl duality and its extensions. Specifically, the structure of the Choi operator for the broadcasting operation, involving both permutation and mixed unitary invariant, leads naturally to the walled Brauer algebra, represented here as the partially transposed permutations matrix algebra $\mathcal{A}_{k,1}^d$. The solution to the SDP must therefore be an element of this algebra, which dramatically constrains the search space and makes an analytical solution possible.
\begin{tcolorbox}
\begin{theorem}\label{thm:broadcasting}
The optimal simulation cost of all universal virtual $k$-broadcasting protocols is given by
    \begin{equation}
        \begin{aligned}
             \gamma_k=
        \log\left(\frac{2kd}{k+d-1}-1\right)
        \end{aligned}
    \end{equation}
    The corresponding optimal protocol is $J^{*}_{A\to B^k}=J^{\cN_1^*}-J^{\cN_2^*}$,
    \begin{equation}\label{eq:solution}
        \begin{aligned}
            J^{\cN_1^*}=&aF_{\sym_k}(\sym_{k-1}),\quad a=\frac{d^2}{d_{\sym_k}},\\
            J^{\cN_2^*}=&bP_{\cS}^{\sym_k},\quad b=\frac{d(k-1)}{d_{\sym_k}(d+k)},
        \end{aligned}
    \end{equation}
    where $d_{\sym_k}:=\tbinom{k+d-1}{k}$ denotes the dimension of the symmetric subspace of $(\mathbb{C}^d)^{\otimes k}$. The projector $F_{\sym_k}(\sym_{k-1})$ labelled by the Young Diagram pair $(\sym_k,\sym_{k-1})$ contained in the ideal $\cM$, while the projector $P_{\cS}^{\sym_k}$ labelled by $\sym_k$ contained in the ideal $\cS$.
\end{theorem}
\end{tcolorbox}
\begin{proof}
    We are going to provide the optimal virtual protocol and the analytical form of the simulation cost, which arises from the mixed unitary symmetry and permutation invariance in the optimal protocol. The proof strategy involves simplifying the original SDP by exploiting the symmetry, and then determining the Choi operator of the optimal protocol with few parameters through the primal and dual formulations.
    
    Firstly, recall the SDP for computing the simulation cost $\gamma_k$, as proposed in Eq.~\eqref{eq:primal_without_rho}. Without loss of generality, we assume the optimal protocol satisfies the symmetry
\begin{equation}\label{eq:symmetry}
    \begin{aligned}
        [J^{*}_{A\to B^k},U^{\otimes k}\otimes \overline{U}]&=0,\quad U\in \textbf{SU}(d),\\
        [J^{*}_{A\to B^k},V(\sigma)\otimes I]&=0,\quad \sigma\in \textbf{S}_k,
    \end{aligned}
\end{equation}
where $V(\sigma)$ denotes the representation of the symmetric group $\textbf{S}_k$, and the permutation operation $V((12\cdots kk+1))$ on total $k+1$ systems has been omitted. The invariant properties in Eq.~\eqref{eq:symmetry} imply that the optimal operator $J^{*}_{A\to B^k}$ belongs to the algebra of the partially transposed permutations $\cA_{k,1}^{d}$. Since the optimal protocol is an HPTP map, it can be decomposed into two CPTP maps, i.e., $J^{*}_{A\to B^k}=J^{\cN_1^*}-J^{\cN_2^*}$. The invariant properties imply the same symmetry for the Choi operators $J^{\cN_1^*}$ and $J^{\cN_2^*}$. As a result, the optimal simulation cost of all universal virtual $k$-broadcasting protocols can be rewritten as the following optimization problem:
\begin{equation}\label{sdp:primal_broadcasting}
\begin{aligned}
    2^{\gamma_k} =  \min\;&p_1+p_2\\
    {\rm s.t.}\; &J^{\cN_1}_{A\to B^k}:=\sum_{\alpha\in \mathbb{Y}_{k-1}^d}\sum_{\mu=\alpha+\yng(1)}a^{(1)}_\mu(\alpha)F_\mu(\alpha)+\sum_{\mu\in\mathbb{Y}_{k}^d}a^{(1)}_\mu P^\mu_{\cS}\\
    &J^{\cN_2}_{A\to B^k}:=\sum_{\alpha\in \mathbb{Y}_{k-1}^d}\sum_{\mu=\alpha+\yng(1)}a^{(2)}_\mu(\alpha)F_\mu(\alpha)+\sum_{\mu\in\mathbb{Y}_{k}^d}a^{(2)}_\mu P^\mu_{\cS}\\
    &\tr_{\backslash AB_1}[J^{\cN_{1}}_{A\to B^k}-J^{\cN_{2}}_{A\to B^k}]=\Psi_{AB_1}^d,\\
    &\tr_{B^k}[J^{\cN_{1}}_{A\to B^k}]=p_1I_{A},\,\tr_{B^k}[J^{\cN_{2}}_{A\to B^k}]=p_2I_{A},\\
    &a^{(j)}_{\mu}(\alpha)\geq 0, a^{(j)}_{\mu}\geq0,\quad j=1,2,
\end{aligned}
\end{equation}
where $\Psi_{AB_1}^d$ is the unnormalized $d\ox d$ maximally entangled state on system $AB_1$. It is a linear programming problem with respect to the optimization variables $a^{(j)}_{\mu}(\alpha)$, $a^{(j)}_{\mu}$, and $p_j$ for $j=1,2$. 

It is noted that the maximally entangled state remains invariant under the action of the symmetric group $\textbf{S}_k$ in the constraint of the above new linear program, which inspires us to investigate the optimal solution within the symmetric subspace. Specifically, we construct a feasible solution as shown in Eq.~\eqref{eq:solution}, such that 
\begin{equation}\label{eq:primal_feasible_solution}
    \begin{aligned}
        J^{*}_{A\to B^k}=\frac{d^2}{d_{\sym_k}}F_{\sym_{k}}(\sym_{k-1})-\frac{d(k-1)}{d_{\sym_{k}}(k+d)}\left(P^{\sym_{k}}\otimes I-F(P^{\sym_k}\otimes I)\right),
    \end{aligned}
\end{equation}
where $F=\sum_{\alpha\in\mathbb{Y}_{k-1}^d}\sum_{\mu=\alpha+\yng(1)}F_{\mu}(\alpha)$ is the identity on the ideal $\cM$, $\sym_k\in\mathbb{Y}_k^d$ and $\sym_{k-1}\in\mathbb{Y}_{k-1}^d$ are Young Diagrams for fully symmetric subspaces. Next, we are going to verify that it is indeed a feasible solution. Due to the constraints on $p_1$ and $p_2$ of the SDP in Eq.~\eqref{sdp:primal_broadcasting}, we have 
\begin{equation}
    \begin{aligned}
        p_1+p_2&=\frac{d\tr[F_{\sym_k}(\sym_{k-1})]}{d_{\sym_k}}+\cdot\frac{k-1}{d_{\sym_{k}}(k+d)}(d_{\sym_k}d-\tr[F_{\sym_k}(\sym_{k-1})])\\
        &=\frac{kd}{k+d-1}+\left(\frac{(k-1)d}{k+d}-\frac{(k-1)k}{(k+d)(k+d-1)}\right)\\
        &=\frac{2kd}{k+d-1}-1,
    \end{aligned}
\end{equation}
where the first equation follows from the fact that $\tr[P^{\mu}]=d_\mu$, $\tr[P^\mu P^\nu]=\delta_{\mu\nu}P^\mu$ for all $\mu,\nu\in\mathbb{Y}_k^d$, which means $F(P^{\sym_k}\otimes I)=F_{\sym_k}(\sym_{k-1})$. The second equation follows from the property $\tr[F_{\mu}(\alpha)]=d_\alpha m_\mu$. For the symmetry subspace labelled by $\sym_k$, we have $m_{\sym_k}=1$. The third equation can be obtained by further simplification.

Furthermore, one can directly check that, by using the auxiliary Lemma~\ref{lem:aux_lemma}, we obtain that the Choi operator $J^{*}_{A\to B^k}$ as shown in Eq.~\eqref{eq:primal_feasible_solution} after undergoing partial trace, is equal to the unnormalized $d\ox d$ maximally entangled state, which satisfies the equation constraint of the SDP in Eq.~\eqref{sdp:primal_broadcasting}. Thus, the operator proposed in Eq.~\eqref{eq:primal_feasible_solution} is a feasible solution, which means that the optimal simulation cost has an upper bound, i.e., $\gamma_k\leq \log\left(\frac{2kd}{k+d-1}-1\right)$. 

Secondly, we focus on the dual SDP proposed in Ref.~\cite{yao2024optimal}, 
\begin{equation}\label{eq:dual_without_rho}
\begin{aligned}
\max\;& \sum_{j=1}^k\tr[X_{AB_j}\Psi_{AB_j}^d]\\
        {\rm s.t.}\;& \tr[Z_{A}]\leq 1, \tr[K_{A}]\leq 1,\\
        & Z_{A}\ox I_{B^k} - \sum_{j=1}^n X_{AB_j}\ox I_{B_2\cdots B_k}\geq 0,\\
        & K_{A}\ox I_{B^k} + \sum_{j=1}^n X_{AB_j}\ox I_{B_2\cdots B_k}\geq 0, \,j=1,\cdots,k,
\end{aligned}
\end{equation}
where $X_{AB_j}$, $Z_A$ and $K_A$ are optimization variables. A general dual SDP for $\eta_k(\rho_{AB})$ can be found in Appendix~\ref{app:dual_sdp}. Then, by showing that $\{X_{AB_1},\cdots,X_{AB_k},Z_{A},K_{A}\}$ is a feasible solution of the dual SDP in Eq.~\eqref{eq:dual_without_rho}, where $Z_{A}=K_{A}=\frac{I_A}{d}$, and $ X_{AB_1}=\cdots=X_{AB_k}=\frac{2}{d(k+d-1)}\Phi^d-\frac{1}{kd}I^{\otimes 2}$, we obtain the lower bound of $\gamma_k$, i.e., $\gamma_k\geq \log\left(\frac{2kd}{k+d-1}-1\right)$ ~\cite{yao2024optimal}. Combining the upper bound, we have $\gamma_k^*=\log\left(\frac{2kd}{k+d-1}-1\right)$, which completes this proof.
\end{proof}
Theorem~\ref{thm:broadcasting} clearly presents the optimal protocol and the associated cost, thereby resolving the open question of what is the optimal virtual $k$-broadcasting protocol left in~\cite{yao2024optimal}. The structure of the optimal solution in Eq~\eqref{eq:solution} is highly insightful. It is a linear combination of two projectors, $F_{\mathrm{sym}_k}(\mathrm{sym}_{k-1})$ and $P_\mathcal{S}^{\mathrm{sym}_k}$, which belong to two orthogonal ideals, $\mathcal{M}$ and $\mathcal{S}$, of the partially transposed permutations algebra. This decomposition is not arbitrary; it reflects a fundamental division in the representation theory of the walled Brauer algebra. Physically, it suggests the optimal non-physical operation is constructed from two distinct types of quantum processes, one residing entirely in the symmetric subspace and related to cloning~\cite{werner1998optimal,Parzygnat2023}, and the other providing a correction.

\begin{shaded}
    \begin{remark}\label{re:iso_expression} Using the same analytical approach, we found that for isotropic states $\Phi_{r}^d:=r\Phi^d+(1-r)\frac{I-\Phi^d}{d^2-1},r\in[0,1]$, it holds that 
        \begin{equation}
            \begin{aligned}
            \eta_k(\Phi_r^d)=\Bigg\{
            \begin{array}{cc}
                0,&  r\in[0,\frac{k+d-1}{kd}]\\
                \log\left(\frac{2kdr}{k+d-1}-1\right), &  r\in(\frac{k+d-1}{kd},1]
            \end{array}.
            \end{aligned}
        \end{equation}
    The case of $\eta_k(\Phi^d_r)=0$ follows from the fact that an isotropic state $\Phi_r^d$ is $k$-extendible if and only if the parameter $r$ belongs to $[0,\frac{k+d-1}{kd}]$~\cite{johnson2013compatible}; for the case of $r\in(\frac{k+d-1}{kd},1]$, one can construct the feasible solutions for the primal and dual SDPs, respectively, as follows: $J^{\ast}_{A\to B^k}=\frac{dr}{d_{\sym_k}}F_{\sym_{k}}(\sym_{k-1})-\frac{rkd-(k+d-1)}{kd-(k+d-1)}\cdot\frac{k-1}{d_{\sym_{k}}(k+d)}\left(P^{\sym_{k}}\otimes I-F(P^{\sym_k}\otimes I)\right)$, and $ X_{AB_1}^\ast=\cdots=X_{AB_k}^\ast=\frac{2d}{k+d-1}\Phi^d-\frac{1}{k}I_{AB}$.
    \end{remark}
\end{shaded}
\subsubsection{Implementation circuits for optimal protocols}\label{sec:quanutm_circuit}
Theorem~\ref{thm:broadcasting} provides the optimal protocol in the form of an abstract operator. 
In what follows we construct the implementation circuits for the optimal virtual broadcasting protocol. In particular, Propositions~\ref{prop:circuit_F} and~\ref{prop:circuit_M} give explicit circuits of the core components for $\mathcal{N}_1^*$ and $\mathcal{N}_2^*$ given in Theorem~\ref{thm:broadcasting}, respectively. This demonstrates the operational feasibility of simulating the optimal virtual protocol.

\begin{figure}[H]
    \centering
    \includegraphics[width=0.8\linewidth]{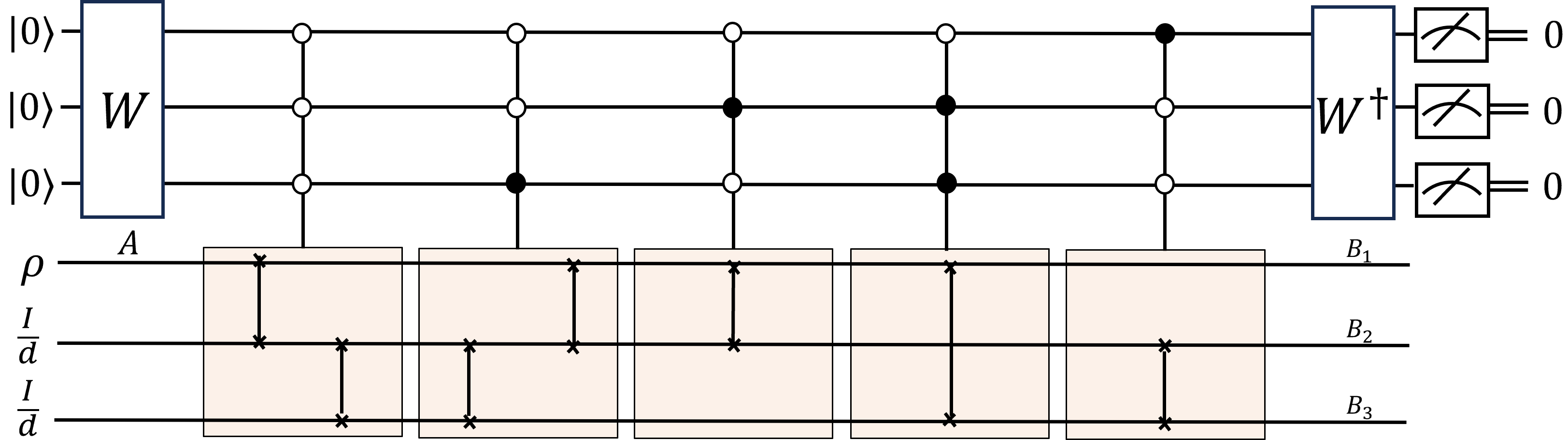}
    \caption{Quantum circuit for implementing the CPTN map $\Lambda_{A\to B_1B_2B_3}$. The preparation circuit $W$ satisfies $W\ket{0}=\sum_{j=0}^{5}\frac{1}{\sqrt{6}}\ket{j}$, and the entire circuit achieve the linear combination of unitaries $V(\sigma)$, for $\sigma\in\textbf{S}_3$. This serves as an example of $k=3$, wherein analogous symmetric projections for $k$ systems can be similarly achieved.}
    \label{fig:sym_proj}
\end{figure}

\begin{tcolorbox}
\begin{proposition}\label{prop:circuit_F}
    The CPTN channel $\Lambda_{A\to B^k}$ with the Choi operator $J_{A\to B^k}^{\Lambda}:=F_{\sym_k}(\sym_{k-1})$ can be implemented using the symmetric projector circuit (an example proposed in Fig.~\ref{fig:sym_proj}), where the operator $F_{\sym_k}(\sym_{k-1})$ is a projector in the ideal $\cM$.
\end{proposition}
\end{tcolorbox}
\begin{proof}
    By definition, the Choi operator can be represented as follows:
    \begin{equation}
        \begin{aligned}
            J_{A\to B^k}^{\Lambda}=(P^{\sym_k}_{AB^{k-1}}\otimes I_{B_{k}}) (I_{AB^{k-2}}\otimes\Psi_{B_{k-1}B_{k}}^d) (P^{\sym_k}_{AB^{k-1}}\otimes I_{B_{k}}),
        \end{aligned}
    \end{equation}
    where $P^{\sym_k}_{AB^{k-1}}:=\frac{1}{k!}\sum_{\sigma\in\textbf{S}_k}V(\sigma)$ denotes the symmetric projector on the first $k$ systems and $\Psi_{B_{k-1}B_{k}}^d$ is the unnormalized $d\ox d$ maximally entangled state on the last two systems. For any given state $\rho\in\mathscr{D}(\cH_{A})$, we have 
    \begin{equation}
        \begin{aligned}
            \Lambda_{A\to B^k}(\rho)&=\tr_{A}[J_{A\to B^k}^\Lambda(\rho_A^T\otimes I_{B^k})]\\
            &=\tr_A[(P^{\sym_k}_{AB^{k-1}}\otimes I_{B_{k}}) (I_{AB^{k-2}}\otimes\Psi_{B_{k-1}B_{k}}^d) (P^{\sym_k}_{AB^{k-1}}\otimes I_{B_{k}})(\rho_A^T\otimes I_{B^k})]\\
            &=\tr_A[(P^{\sym_k}_{AB^{k-1}}\otimes I_{B_{k}}) (I_{B^{k-1}}\otimes\Psi_{AB_{k}}^d) (P^{\sym_k}_{AB^{k-1}}\otimes I_{B_{k}})(\rho_A^T\otimes I_{B^k})]\\
            &=\frac{1}{d^{k-1}}\tr_A[\tr_{A^{k-1}}[(P^{\sym_k}_{AB^{k-1}}\otimes I_{B_kA^{k-1}})(\Psi_{AB^{k-1}B_kA^{k-1}}^d)(P^{\sym_k}_{AB^{k-1}}\otimes I_{B_kA^{k-1}})](\rho_A^T\otimes I_{B^k})]\\
            &=\frac{1}{d^{k-1}}\tr_{AA^{k-1}}[(I_{AB^{k-1}}\otimes P^{\sym_k}_{B_kA^{k-1}}) (\Psi_{AB^{k-1}B_kA^{k-1}}^d) (I_{AB^{k-1}}\otimes P^{\sym_k}_{B_kA^{k-1}}) (\rho_A^T\otimes I_{B^k}\otimes I_{A^{k-1}})]\\
            &=\frac{1}{d^{k-1}}\tr_{AB^{k-1}}[(I_{AB^{k-1}}\otimes P^{\sym_k}_{B_kA^{k-1}}) (\Psi_{AB^{k-1}B_kA^{k-1}}^d) (I_{AB^{k-1}}\otimes P^{\sym_k}_{B_kA^{k-1}}) (\rho_A^T\otimes I_{B^k}\otimes I_{A^{k-1}})]\\
            &=\tr_{AB^{k-1}}[J_{AB^{k-1}\to B_kA^{k-1}}^{\cP_k}(\rho_A^T\otimes \frac{I_{B^{k-1}}}{d^{k-1}}\otimes I_{B_kA^{k-1}})],
        \end{aligned}
    \end{equation}
    where the third equation follows from the fact that $P^{\sym_k}V(\sigma)=P^{\sym_k}$ for all permutation operators $V(\sigma)$, $\sigma\in\textbf{S}_k$, the fifth and sixth inequalites follow from the property that $Y\otimes Z\dket{X}=\dket{YXZ^T}$~\cite{hayashi2017group} ($\dketbra{I}{I}=\Psi$) and the symmetry of the Young projector $P^{\sym_k}$. Note that the notation $J_{AB^{k-1}\to B_kA^{k-1}}^{\cP_k}$ is the Choi operator of the symmetry map $\cP_{{AB^{k-1}\to B_kA^{k-1}}}(\cdot):=P^{\sym_k}(\cdot)P^{\sym_k}$, which project any input state in the space $\cH_{A}\otimes \cH_{B^{k-1}}$ onto its symmetric sub-space.

    Finally, based on the Ref.~\cite{yao2025protocols}, one can employ the block-encoding technology~\cite{low2019hamiltonian} to implement this symmetry map. An example can be found in Fig.~\ref{fig:sym_proj}, which completes this proof.
\end{proof}

\begin{tcolorbox}
    \begin{proposition}\label{prop:circuit_M}
        The CPTN channel $\Gamma_{A\to B^{k-1}}$ with the Choi operator $J_{A\to B^{k-1}}^{\Gamma}:=P^{\sym_k}$ can be implemented by tracing out the system $B_k$ of the realization circuit of $J_{A\to B^k}^{\Lambda}$.
        
    \end{proposition}
\end{tcolorbox}
\begin{proof}
    By tracing out the last system of $J_{A\to B^k}^{\Lambda}$, we have
    \begin{equation}
    \begin{aligned}
        \tr_{B_k}[J_{A\to B^k}^\Lambda]
        &=\tr_{B_k}[F_{\sym_{k}}(\sym_{k-1})]\\
        &=\frac{d_{\sym_{k-1}} }{kd_{\sym_{k}}} P^{\sym_{k}}\sum_{j=1}^{k}S_{j,k}P^{\sym_{k-1}} S_{j,k} P^{\sym_{k}}\\
        &=\frac{d_{\sym_{k-1}}}{d_{\sym_{k}}} P^{\sym_{k}},
    \end{aligned}
\end{equation}
where $d_{\sym_k}:=\tbinom{k+d-1}{k}$ denotes the dimension of the symmetric subspace of $(\mathbb{C}^d)^{\otimes k}$, and the second equation follows from the fact that $P^{\sym_k}S_{j,k}=P^{\sym_k}$ for all $j=1,\cdots,k$ and $P^{\sym_k} P^{\sym_{k-1}}=P^{\sym_k}$. Therefore, one can obtain the quantum circuit by tracing out the last system of the implementation circuit, as shown in Fig.~\ref{fig:sym_proj} (example for $k=3$), which completes this proof.
\end{proof}
\subsection{Quantifying unextendibility of entanglement}

This subsection aims to establish the virtual extension cost $\eta_k(\rho_{AB})$ as a well-behaved and meaningful entanglement measure for a general bipartite state $\rho_{AB}$. We will do so by connecting it to the established framework of quantum resource theories and comparing its properties to other known entanglement measures.
\subsubsection{Robustness of $k$-unextendibility}
We begin by introducing the robustness of unextendibility, a measure defined within the axiomatic framework of resource theories. Specifically, we focus on the absolute robustness $\cR_{\rm{absolute}}^{{\rm EXT}_k}(\rho)$~\cite{gour2024resourcesquantumworld,harrow2003robustness}, which quantifies how much $k$-extendible "noise" must be mixed with a state $\rho_{AB}$ to render it $k$-extendible.
Theorem~\ref{thm:neg_equal_robust} reveals a deep and precise mathematical equivalence between our operational cost $\eta_k(\rho_{AB})$ and this axiomatic robustness measure.

\begin{tcolorbox}
\begin{theorem}[Robustness]\label{thm:neg_equal_robust}
    The following relation holds for any state $\rho_{AB}$:
    \begin{align}
        \eta_k(\rho_{AB})  = \log(2\cR_{\rm{absolute}}^{{\rm EXT}_k}(\rho_{AB}) + 1),
    \end{align}
    where
    $\cR_{\rm{absolute}}^{{\rm EXT}_k}(\rho_{AB})$ is the absolute robustness of $k$-extendibility (Def.~\eqref{def:rob}).
\end{theorem}
\end{tcolorbox}
\begin{proof}
    According to Eq.~\eqref{eq:cost_as_state_opt},
    the
    virtual state $k$-extension cost
    can be characterized by the following SDP:
    \begin{align}\label{eq:k_ext_sdp}
        2^{\eta_k(\rho_{AB})} = \min \tr[N] ~\text{s.t.}~ -N \leq M_{AB^k} \leq N,~ M_{AB_j} = \rho_{AB} \,\forall\, 1\leq j \leq k,
    \end{align}
    and the robustness can be characterized as
    \begin{align}
        &2\cR_{\rm{absolute}}^{{\rm EXT}_k}(\rho_{AB}) + 1 \nonumber\\
        =& \min \tr[M^+_{AB^k} + M^-_{AB^k}] ~\text{s.t.}~ 0\leq M^\pm_{AB^k},~ M^\pm_{AB_1} = M^\pm_{AB_j} \,\forall\, 2\leq j \leq k,~ M^+_{AB_1} - M^-_{AB_1} = \rho_{AB}.\label{eq:robustness_sdp}
    \end{align}
    It is clear that $2\cR_{\rm{absolute}}^{{\rm EXT}_k}(\rho_{AB}) + 1 \geq 2^{\eta_k(\rho_{AB})}$ because every pair of feasible $M^\pm_{AB^k}$ provides a feasible $M_{AB^k} \coloneqq M^+_{AB^k} - M^-_{AB^k}$ and $\tr[M^+_{AB^k} + M^-_{AB^k}] \geq \|M_{AB^k}\|_1$.

    Now, suppose that $M_{AB^k}$ is an optimal solution to SDP~\eqref{eq:k_ext_sdp}, whose positive part and negative part are denoted by $M^+_{AB^k}$ and $M^-_{AB^k}$, respectively. Define
    \begin{align}
        \widehat{M}^\pm_{AB^k} \coloneqq \frac{1}{k!}\sum_{\pi\in \textbf{S}_k} V(\pi)\left(M^\pm_{AB^k}\right)V(\pi)^\dagger,
    \end{align}
    where $\textbf{S}_k$ is the symmetric group of degree $k$ and $V(\pi)$, acting on $B^k$, is the permutation operator associated with $\pi$.
    It can be verified that $\widehat{M}^\pm_{AB_1} = \widehat{M}^\pm_{AB_j}$ for every $2\leq j\leq k$. Furthermore,
    \begin{align}
        \widehat{M}^+_{AB^k} - \widehat{M}^-_{AB^k} = \frac{1}{k!}\sum_{\pi\in \textbf{S}_k} V(\pi)\left(M^+_{AB^k} - M^-_{AB^k}\right)V(\pi)^\dagger
    \end{align}
    implies that $\widehat{M}^+_{AB_1} - \widehat{M}^-_{AB_1} = \rho_{AB}$ since $\tr_{B_2\cdots B_k}[V(\pi)(M_{AB^k})V(\pi)^\dagger] = \rho_{AB}$ due to the last constraint in SDP~\eqref{eq:k_ext_sdp}.
    Thus, $\widehat{M}^+_{AB^k}$ and $\widehat{M}^-_{AB^k}$ form a feasible solution to SDP~\eqref{eq:robustness_sdp}. Because
    \begin{align}
        \tr\left[\widehat{M}^+_{AB^k} + \widehat{M}^-_{AB^k}\right] &= \frac{1}{k!}\sum_{\pi\in \textbf{S}_k} \tr\left[V(\pi)\left(M^+_{AB^k} + M^-_{AB^k}\right)V(\pi)^\dagger\right]\\
        &= \frac{1}{k!}\sum_{\pi\in \textbf{S}_k} \tr\left[M^+_{AB^k} + M^-_{AB^k}\right]\\
        &= \tr\left[M^+_{AB^k} + M^-_{AB^k}\right]\\
        &= \left\|M_{AB^k}\right\|_1\\
        &= 2^{\eta_k(\rho_{AB})},
    \end{align}
    which implies $2\cR_{\rm{absolute}}^{{\rm EXT}_k}(\rho_{AB}) + 1 \leq 2^{\eta_k(\rho_{AB})}$.
    Therefore, we conclude that $2^{\eta_k(\rho_{AB})} = 2\cR_{\rm{absolute}}^{{\rm EXT}_k}(\rho_{AB}) + 1$.
\end{proof}

This equivalence demonstrates that our operational measure, defined via the cost of a virtual task, is mathematically identical to the absolute robustness, which is defined via mixing with free states. This endows the abstract notion of robustness with a concrete, physical interpretation. 

It is also important to distinguish our measure from the global robustness of extendibility studied in \cite{kaur2021resource}, which allows mixing with arbitrary states.
Specifically, consider states of the form $\rho_r:=(\ketbra{\psi_0}{\psi_0} + \ketbra{\psi_1}{\psi_1})/2$, where $\ket{\psi_0}:=\sqrt{1-r}\ket{00}+\sqrt{r}(\ket{01}+\ket{12})/\sqrt{2}$ and $\ket{\psi_1}:=\sqrt{1-r}\ket{11}+\sqrt{r}(\ket{12}+\ket{20})/\sqrt{2}$. When $r=r_0=0.65$, we numerically found that $\cR_{\rm{global}}^{\rm EXT_2}(\rho_{r_0})<0.20567460<0.20603501<\cR_{\rm{absolute}}^{\rm EXT_2}(\rho_{r_0})$, yielding a gap between the two unextendibility measure.
\begin{shaded}
\begin{remark}\label{thm:normalization} Combining Theorem~\ref{thm:neg_equal_robust} and Remark~\ref{re:iso_expression},
we have 
\begin{equation}
    \begin{aligned}
    \cR_{\rm{absolute}}^{{\rm EXT}_k}(\Phi_r^d)=\Bigg\{
    \begin{array}{cc}
        1,&  r\in[0,\frac{k+d-1}{kd}]\\
        \frac{kdr}{k+d-1}-1, &  r\in(\frac{k+d-1}{kd},1]
    \end{array},
    \end{aligned}\label{eq:iso_rob}
\end{equation}
where $\Phi_{r}^d$ denotes the isotropic state with parameter $r\in[0,1]$. We remark that \cite{email} independently developed Theorem~\ref{thm:neg_equal_robust} and derived Eq. \eqref{eq:iso_rob} from the definition of robustness and the $k$-extendibility. However, this approach does not yield further information regarding the optimal protocol, as well as its circuit implementation. 
\end{remark}
\end{shaded}
\subsubsection{Exact entanglement distillation}
To further explore the properties of $\eta_k(\rho_{AB})$ as a useful entanglement measure, we first examine its implications for a fundamental entanglement manipulation task: one-shot exact entanglement distillation.

The one-shot 1-LOCC exact distillable entanglement of $\rho_{AB}$ is defined as
\begin{align}
    E^{(1)}_{1-LOCC}(\rho_{AB})
    :=\sup \{\log d:\cN_{AB \to \hat{A}\hat{B}}(\rho_{AB})=\Phi^d_{\hat{A}\hat{B}},\cN_{AB \to \hat{A}\hat{B}}\in 1-LOCC\},
\end{align}
where $\Phi^d_{\hat{A}\hat{B}}$ is $d\otimes d$ maximally entangled state.
\begin{tcolorbox}
\begin{proposition}
    For any bipartite state $\rho_{AB}\in\mathscr{D}(\cH_A\otimes \cH_B)$, its one-shot 1-LOCC exact distillable entanglement has the following upper bound:
    \begin{align}
        E^{(1)}_{1-LOCC}(\rho_{AB}) \leq  \min_{k}\,\log \left(\frac{(k-1)(\cR_{\rm{absolute}}^{{\rm EXT}_k}(\rho_{AB})+1)}{k-1-\cR_{\rm{absolute}}^{{\rm EXT}_k}(\rho_{AB})}\right).
    \end{align}
\end{proposition}
\end{tcolorbox}
\begin{proof}
For a given bipartite state $\rho_{AB}$, suppose there exists a 1-LOCC channel $\cM_{AB\to\hat{A}\hat{B}}$ that transforms $\rho_{AB}$ to the maximally entangled state $\Phi^d_{\hat{A}\hat{B}}$ exactly. Then, by the monotonicity of this robustness $\cR_{\rm{absolute}}^{{\rm EXT}_k}$, the following inequality holds
\begin{align}\label{eq:ub_by_monotonicity}
    \cR_{\rm{absolute}}^{{\rm EXT}_k}(\rho_{AB}) \geq \cR_{\rm{absolute}}^{{\rm EXT}_k}(\cM_{AB\to\hat{A}\hat{B}}(\rho_{AB}))=\cR_{\rm{absolute}}^{{\rm EXT}_k}(\Phi^d_{\hat{A}\hat{B}})=\frac{kd}{k+d-1}-1,
\end{align}
where the last equation follows from Theorem~\ref{thm:normalization}. Then, by optimizing over all 1-LOCC channels $\cN_{AB \to \hat{A}\hat{B}}$ and natural number $k\in\NN_+$, we have
\begin{equation}
    \begin{aligned}
        \frac{(k-1)(\cR_{\rm{absolute}}^{{\rm EXT}_k}(\rho_{AB})+1)}{k-1-\cR_{\rm{absolute}}^{{\rm EXT}_k}(\rho_{AB})} \geq d,
    \end{aligned}
\end{equation}
which completes this proof.
\end{proof}
\subsubsection{Comparison with logarithmic negativity}

A crucial benchmark for any new entanglement measure is its relationship to established, well-understood quantifiers. Here, we compare our virtual extension cost to the logarithmic negativity, one of the most widely used measures of entanglement. It is expected that in the limit of infinite extensions ($k \to \infty$), where the set of free states ($k$-extendible states) approaches the set of all separable states, a measure of unextendibility should converge to a standard entanglement monotone. Proposition~\ref{prop:comparison_EN} confirms this intuition for pure states, showing that our measure indeed converges to the logarithmic negativity. This provides a strong consistency check and grounds our operational measure within the broader landscape of entanglement theory.

\begin{tcolorbox}
    \begin{proposition}\label{prop:comparison_EN}
    For any bipartite pure state ${\psi}_{AB}\in\mathscr{D}(\cH_{A}\otimes \cH_B)$, we have
        \begin{equation}
            \begin{aligned}
                 \lim_{k\to \infty} \log\left(\frac{2^{\eta_k(\psi_{AB})}+1}{2}\right) = E_N(\psi_{AB}),
            \end{aligned}
        \end{equation}
    where $E_N(\psi_{AB}):=\log||\psi_{AB}^{T_B}||_1$ denotes the logarithmic negativity of $\psi_{AB}$.
    \end{proposition}
\end{tcolorbox}
\begin{proof}
    By definition of $\eta_k(\cdot)$ and Theorem~\ref{thm:neg_equal_robust}, for all $k$, we have
    \begin{equation}
        \begin{aligned}
            \log\left(\frac{2^{\eta_k(\psi_{AB})}+1}{2}\right)=\log(\cR_{\rm{absolute}}^{{\rm EXT}_k}(\psi_{AB})+1).
        \end{aligned}
    \end{equation}
    As $k\to \infty$, the $k$-extendible state converges to the separable state, which means $\cR_{\rm{absolute}}^{\rm SEP}=\cR_{\rm{absolute}}^{\rm \EXT_\infty}$. Recall the definition of global robustness, and the properties of the max-relative entropy $E_{\text{max}}$ of entanglement and logarithmic negativity, we have 
    \begin{equation}
        \begin{aligned}
            E_N(\psi_{AB}) = E_{\text{max}}(\psi_{AB})=\log(\cR_{\rm{global}}^{\rm SEP}(\psi_{AB})+1)=\log(\cR^{\rm SEP}_{\rm{absolute}}(\psi_{AB})+1),
        \end{aligned}
    \end{equation}
    where the first two equations follow from Ref.~\cite{datta2009max}, and the third equation follows from Ref.~\cite{steiner2003generalized,harrow2003robustness}, which completes this proof.
\end{proof}
\section{Conclusions and discussions}\label{sec:conclusion}
In this paper, we introduced a novel, operationally-driven perspective to quantify the unextendibility of entanglement by defining the virtual extension cost, $\eta_k(\cdot)$, as a new entanglement measure.
First, we established a fundamental link between this measure and quantum communication, proving that the virtual extension cost of a maximally entangled state is identical to the optimal cost of universal virtual quantum broadcasting. Second, using the representation theory of the walled Brauer algebra, we derived the first analytical, closed-form expression for this optimal broadcasting cost and gave the explicit algebraic form of the optimal protocol, resolving an open question from \cite{yao2024optimal}. Third, we bridged the operational and axiomatic viewpoints by proving that the virtual extension cost is mathematically equivalent to the absolute robustness of $k$-unextendibility, thereby endowing the latter with a clear physical interpretation. Finally, we demonstrated the well-behaved nature of $\eta_k(\cdot)$ as a bona fide entanglement measure by deriving bounds on entanglement distillation and showing its convergence to logarithmic negativity in the appropriate limit.

Several open questions and future directions emerge from our work. A natural next step is to solve the virtual extension cost for other important classes of states, such as Werner states or general pure states. The multipartite generalization of virtual state extension presents another intriguing avenue, potentially connecting to measures of genuine multipartite entanglement. Furthermore, exploring the application of the measure we proposed for private key distillation and private communication over quantum channels is also valuable.
Finally, a deeper investigation into the relationship between $\eta_k(\cdot)$ and other entanglement measures for general mixed states could further elucidate the structure of entanglement from the perspective of its non-shareability.

\section{Acknowledgment}
We would like to thank Benchi Zhao for his helpful comments.
H.Y. would like to thank Qian Chen and Yunlong Xiao for helpful discussions on the representation theory during the AQIS Conference.
X.Z. and X.W. would like to thank Ludovico Lami and Mark Wilde for private communications on the virtual extension cost and its usefulness in quantifying entanglement.
This work was partially supported by the National Key R\&D Program of China (Grant No.~2024YFE0102500), the National Natural Science Foundation of China (Grant. No.~12447107), the Guangdong Provincial Quantum Science Strategic Initiative (Grant No.~GDZX2403008, GDZX2403001), the Guangdong Provincial Key Lab of Integrated Communication, Sensing and Computation for Ubiquitous Internet of Things (Grant No.~2023B1212010007), the Quantum Science Center of Guangdong-Hong Kong-Macao Greater Bay Area, and the Education Bureau of Guangzhou Municipality.
\bibliographystyle{alpha}
\bibliography{ref}

\appendix
\setcounter{subsection}{0}
\setcounter{table}{0}
\setcounter{figure}{0}

\vspace{3cm}
\onecolumngrid
\newpage
\begin{center}
\Large{\textbf{Appendix for} \\ \textbf{Quantifying Unextendibility via Virtual State Extension
}}
\end{center}

\renewcommand{\theequation}{S\arabic{equation}}
\renewcommand{\thesubsection}{\normalsize{ \arabic{subsection}}}
\renewcommand{\theproposition}{S\arabic{proposition}}
\renewcommand{\thedefinition}{S\arabic{definition}}
\renewcommand{\thefigure}{S\arabic{figure}}
\setcounter{equation}{0}
\setcounter{table}{0}
\setcounter{section}{0}
\setcounter{proposition}{0}
\setcounter{definition}{0}
\setcounter{figure}{0}
\section{Proof of Auxiliary Lemmas}\label{app:aux_lemma}
\begin{shaded}
\begin{lemma}\label{lem:recursive_lemma} For any $d\times d$ matrix $X\in M_d(\mathbb{C})$, we denote $R_n(X):=\tr_{2\cdots n}[P^{\sym_n}(I^{\otimes n-1}\otimes X)P^{\sym_n}]$, where $P^{\sym_n} := \frac{1}{n!} \sum_{\pi \in S_n} V(\pi)$, and $V(\pi)$ is the permutation representation of $\pi\in\textbf{S}_n$ on the Hilbert space $(\mathbb{C}^d)^{\otimes n}$. Then for \( n \ge 2 \), we have
\begin{equation}\label{eq:target_eq}
    R_n(X) = a_n\tr(X)I + b_nX,
\end{equation}
where $a_n := \frac{(n-1)d_{\sym_n}}{dn(d+1)}$, $b_n :=\frac{(d+n)d_{\sym_n}}{dn(d+1)}$, $d_{\sym_n}:=\tbinom{n+d-1}{n}$ denotes the dimension of the symmetric subspace of $(\mathbb{C}^d)^{\otimes n}$. and $I$ is the identity operator on space $\mathbb{C}^d$.
\end{lemma}
\end{shaded}
\begin{proof}
By definition and the properties of permutation operators, we have,
\begin{equation}\label{extension}
    \begin{aligned}
        R_n(X) &= \frac{1}{(n!)^2} 
        \sum_{\pi, \sigma \in \textbf{S}_n} 
        \tr_{2 \cdots n}\bigl[
        V(\pi)\bigl(I^{\otimes (n-1)} \otimes X\bigr)V(\sigma)
        \bigr]\\
        &=\frac{1}{(n!)^2} 
        \sum_{\pi, \sigma \in \textbf{S}_n} 
        \tr_{2 \cdots n}\bigl[
        V(\pi\sigma)V(\sigma^{-1})(I^{\otimes (n-1)} \otimes X)V(\sigma)
        \bigr]\\
        &\overset{\tau:=\pi\sigma}{=}\frac{1}{(n!)^2} \sum_{\tau\in\textbf{S}_n}
        \sum_{\sigma \in \textbf{S}_n} 
        \tr_{2 \cdots n}\bigl[
        V(\tau)E_{\sigma(n)}
        \bigr],
    \end{aligned}
\end{equation}
where the second equation follows from the fact that $V(\pi) = V(\pi\sigma)V(\sigma^{-1})$, while for the third equation derives from the group symmetry and the redefinition, $E_{\sigma(n)}:=I^{\otimes (k-1)} \otimes X \otimes I^{\otimes (n-k)}$ for $\sigma(n)=k$. Note that only the positional variation of the $n$-th system impacts the overall value. It is straightforward to see that the number of permutations mapping the $n$-th system to the $k$-th system is equal to $(n-1)!$, since the remaining $n-1$ elements can be permuted arbitrarily, i.e.,
\begin{equation}
    \#\{\sigma \in \textbf{S}_n : \sigma(n)=k\} = (n-1)!.
\end{equation}
As a result, we obtain
\begin{equation}\label{count_change}
    R_n(X) = \frac{1}{n\,n!} 
\sum_{\tau \in \textbf{S}_n} 
\sum_{k=1}^n 
\tr_{2 \cdots n}\bigl[V(\tau) E_k\bigr].
\end{equation}
Next, we rewrite the permutation $\tau\in\textbf{S}_n$ as a product of disjoint cycles. Let $c(\tau)$ denote the number of cycles, and let $c_0$ be the cycle containing the $1$-th system. Then for any operator $A_j\in M_d(\mathbb{C})$, $j=1,\cdots,n$, we have the following fact that
\begin{equation}\label{tensor_trace}
    \tr_{2 \cdots n}[V(\tau)(A_1 \otimes \cdots \otimes A_n)]=
\Bigl(\prod_{\substack{c\neq c_0 \\ \text{cycles}}} \operatorname{Tr}\bigl(\prod_{i\in c} A_i\bigr)\Bigr)
\Bigl(\prod_{i\in c_0} A_i\Bigr).
\end{equation}
In this work, we only consider the scenario where $A_j = I$ for $j\neq k$ and $A_k = X$, which implies two distinct cases to be examined: the situation where $k$ and $1$ reside within the same cycle, and the alternative scenario where they do not share the same cycle. By definition of operator $A_j$ and the fact in Eq.~\eqref{tensor_trace}, we have the following calculation:
\begin{equation}
    \tr_{2 \cdots n}\bigl[V(\tau) E_k\bigr]=
    \begin{cases}
         d^{c(\tau)-1} X, & \text{if } k \text{ and } 1 \text{ belong to the same cycle };\\
         d^{c(\tau)-2} \tr(X) I, &\text{otherwise.}
    \end{cases}
\end{equation}
Substitute the above result back into the \eqref{count_change}, we have
\begin{equation}
    R_n(X)
= \frac{1}{n\,n!} \sum_{\tau \in \textbf{S}_n} \Biggl[
\sum_{k : k \sim_\tau 1} d^{c(\tau)-1} X
+ \sum_{k : k \not\sim_\tau 1} d^{c(\tau)-2} \operatorname{Tr}(X) I
\Biggr],
\end{equation}
where the notation $k \sim_\tau 1$ means that the system $k$ belongs to the same cycle as the system $1$ in the permutation $\tau$. Furthermore, for a fixed $\tau\in\textbf{S}_{n}$, let $L(\tau)$ denote the length of the cycle containing the $1$-th system. 
Since the operator $X$ applying on different systems $k$ within the same cycle performs identical functions, which implies that the first inner sum contains $L(\tau)$ terms, and the second has $n-L(\tau)$ terms. i.e.,
\begin{equation}
    R_n(X)
= \frac{1}{n\,n!} \sum_{\tau \in \textbf{S}_n} \Bigl[
L(\tau)\, d^{c(\tau)-1} X + (n-L(\tau))\, d^{c(\tau)-2} \tr(X) I
\Bigr].
\end{equation}
Next, we investigate the following sub-term
\begin{equation}
    \begin{aligned}
    W(n) :=& \sum_{\tau \in \textbf{S}_n} L(\tau) d^{c(\tau)}\\=&\sum_{\tau \in S_n} \sum_{j=1}^n \delta_{ j \sim 1 } d^{c(\tau)}\\
    =&K(n)+(n-1)Q(n),
    \end{aligned}
\end{equation}
where $\delta_{j\sim1}$ is the indicator function, $\delta_{j\sim1}=1$ if the system $j$ belongs to the same cycle as $1$, otherwise zero. Separate the cases $j=1$ and $j \neq 1$ and denote $K(n) := \sum_{\tau \in S_n} d^{c(\tau)}$, and $Q(n) := \sum_{\tau \in S_n,j\neq 1,j \sim 1} d^{c(\tau)}$. The third equation follows from the symmetry property, i.e., for any $j \neq 1$, $Q(n)$ has the same value. Classifying all permutations in $S_n$ according to the number of cycles $k$ and applying the key identity of the (unsigned) Stirling numbers of the first kind, we obtain:
\begin{equation}
    K(n) = \sum_{k=1}^n \tbinom{n}{k} d^k = d(d+1)\cdots(d+n-1),
\end{equation}
which also gives the recurrence relation:
\begin{equation}\label{eq:K_recursion}
    K(n) = (d+n-1) K(n-1), \quad K(1) = d.
\end{equation}
When constructing all $\tau \in S_n$ from $\tau' \in S_{n-1}$, the relation between a fixed $j \ne 1$ and $1$ is preserved: if $j \sim 1$ in $\tau'$, then $j \sim 1$ in every $\tau$, while if $j \not\sim 1$ in $\tau'$, then $j \not\sim 1$ in every $\tau$. Hence the contributions to $V(n)$ come only from those $\tau' \in S_{n-1}$ with $j \sim 1$, and for each such $\tau'$ the insertion of $n$ as a new cycle multiplies the weight by $d$, while insertion into any of the $n-1$ positions in existing cycles keeps the weight unchanged, so overall each $\tau'$ with $j \sim 1$ is scaled by the factor $d+n-1$ when generating $S_n$. Thus, we obtain the recurrence
\begin{equation}\label{eq:Q_recursion}
   Q(n) = (d+n-1)Q(n-1), \quad \text{for } n \geq 3,\quad Q(2)=d.
\end{equation}
By combaining Eq.~\eqref{eq:K_recursion} and Eq.~\eqref{eq:Q_recursion}, one can obtain
\begin{equation}
    W(n) = K(n)\frac{d+n}{d+1}.
\end{equation}
As a result, we have
\begin{equation}
    b_n = \frac{1}{dnn!} W(n) = \frac{1}{dnn!} K(n) \frac{d+n}{d+1}=\frac{(d+n)d_{\sym_n}}{dn(d+1)},
\end{equation}
where $d_{\sym_n}=\tbinom{n+d-1}{n}= \frac{K(n)}{n!}$. Similarly, let
$W'(n) := \sum_{\tau}(n-L(\tau))\,d^{\,c(\tau)}$. It is straightforward to see that $W'(n) = nK(n) - W(n)$, which implies the following derivation:
\begin{equation}
    W'(n) = nK(n) - K(n)\frac{d+n}{d+1}
= K(n)\frac{(n-1)d}{d+1}.
\end{equation}
Hence
\begin{equation}
    a_n
= \frac{1}{n\,n!}\cdot\frac{1}{d^2}\cdot K(n)\frac{(n-1)d}{d+1} = \frac{(n-1)d_{\sym_n}}{dn(d+1)},
\end{equation}
which completes this proof.
\end{proof}

\begin{shaded}
\begin{lemma}\label{lem:aux_lemma}
    \begin{equation}
        \begin{aligned}
             \tr_{2\cdots n}[P^{\sym_n}]&=\frac{d_{\sym_n}}{d}I,\\
             \tr_{2\cdots n}[F_{\sym_n}(\sym_{n-1})]&=c_1 I^{\otimes 2} + c_2\Psi,
        \end{aligned}
    \end{equation}
    where $c_1=\frac{(n-1)d_{\sym_{n-1}}}{dn(d+1)}$ and $c_2=\frac{(d+n)d_{\sym_{n-1}}}{dn(d+1)}$, $d_{\sym_n}:=\tbinom{n+d-1}{n}$ denotes the dimension of the symmetric subspace of $(\mathbb{C}^d)^{\otimes n}$. $\Psi:=\sum_{i,j}\ketbra{ii}{jj}$ and $I$ are unnormalized $d\ox d$ maximally entangled state and identity operator, respectively.
\end{lemma}
\end{shaded}
\begin{proof}
    It is straightforward to check the first equation. We mainly focus on the second equation, i.e.,
    \begin{align}\label{IPhi}
        \begin{aligned}
            \tr_{2\cdots n}[F_{\sym_n}(\sym_{n-1})] &=\frac{d_{\sym_{n-1}}}{d_{\sym_n}}\tr_{2\cdots n}[(P^{\sym_n}\otimes I)(I^{\otimes n-1}\otimes \Phi)(P^{\sym_n}\otimes I)]\\
            &=\frac{d_{\sym_{n-1}}}{d_{\sym_n}}\tr_{2\cdots n}[(P^{\sym_n}\otimes I)(I^{\otimes n-1}\otimes \sum_{i,j}\ketbra{ii}{jj})(P^{\sym_n}\otimes I)]\\
            &=\frac{d_{\sym_{n-1}}}{d_{\sym_n}}\sum_{i,j}\tr_{2\cdots n}[P^{\sym_n}(I^{\otimes n-1}\otimes \ketbra{i}{j})P^{\sym_n}] \otimes \ketbra{i}{j},
        \end{aligned}
    \end{align}
    where $P^{\sym_n}:=\frac{1}{n!}\sum_{\sigma\in \textbf{S}_n}V(\sigma)$ denotes the Young projector on the symmetric subspace labelled by the Young Diagram $\sym_n$.
    According to Lemma \ref{lem:recursive_lemma}, let $X=\ketbra{i}{j}$, we obtain $\tr_{2\cdots n}\left[P^{\sym_n}(I^{\otimes n-1}\otimes \ketbra{i}{j})P^{\sym_n}\right]=\frac{(n-1)d_{\sym_n}}{dn(d+1)}\delta_{ij} I + \frac{(d+n)d_{\sym_n}}{dn(d+1)} \ketbra{i}{j}.$
Inserting it into Eq.~\eqref{IPhi} yields
\begin{align}\label{IPhi_substitute}
        \begin{aligned}
            \tr_{2\cdots n}[F_{\sym_n}(\sym_{n-1})] &=\frac{d_{\sym_{n-1}}}{d_{\sym_n}}\sum_{i,j}\left(\frac{(n-1)d_{\sym_n}}{dn(d+1)}\delta_{ij} I + \frac{(d+n)d_{\sym_n}}{dn(d+1)} \ketbra{i}{j}\right) \otimes \ketbra{i}{j}\\
            &=\sum_{i,j}\frac{(n-1)d_{\sym_{n-1}}}{dn(d+1)}\delta_{ij} I \otimes \ketbra{i}{j}+ \sum_{i,j}\frac{(d+n)d_{\sym_{n-1}}}{dn(d+1)} \ketbra{i}{j} \otimes \ketbra{i}{j}\\
            &=\frac{(n-1)d_{\sym_{n-1}}}{dn(d+1)} I^{\otimes 2} + \frac{(d+n)d_{\sym_{n-1}}}{dn(d+1)} \Psi.
        \end{aligned}
    \end{align}
    which completes this proof.
\end{proof}

\begin{remark}
  $\tr_{2\cdots n}\left[P^{\sym_n}(I^{\otimes n-1}\otimes \ketbra{i}{j})P^{\sym_n}\right]$ can also be calculated directly using the tensor network tools. We take $n=3$ as an example to help understand the proof process of Lemma \ref{lem:recursive_lemma} and verify the correctness of the results. The sub-terms $\tr_{23}[V(\sigma)(I^{\otimes 2}\otimes\ketbra{i}{j})V(\tau)]$ are as shown in the table~\ref{fig:tn_contraction}. According to table~\ref{fig:tn_contraction}, we know 
\begin{equation}\label{subterms}
        \begin{aligned}
            \tr_{23}[P^{\sym_3}(I^{\otimes 2}\otimes \ketbra{i}{j})P^{\sym_3}]
            &=\tr_{23}\left[\frac{1}{3!}\sum_{\sigma\in \textbf{S}_n}V(\sigma)(I^{\otimes 2}\otimes \ketbra{i}{j})\frac{1}{3!}\sum_{\tau\in \textbf{S}_n}V(\tau)\right]\\
            &=\frac{1}{36}\sum_{\sigma,\tau\in \textbf{S}_n}\tr_{23}\left[V(\sigma)(I^{\otimes 2}\otimes \ketbra{i}{j})V(\tau)\right]\\
            &=\frac{1}{36}(4d\delta_{ij}I+8\delta_{ij}I+2d^2\ketbra{i}{j}+10d\ketbra{i}{j}+12\ketbra{i}{j})\\
            &=\frac{(d+2)}{9}\delta_{ij}I+\frac{(d+2)(d+3)}{18}\ketbra{i}{j},
        \end{aligned}
    \end{equation}
which is consistent with results in Lemma~\ref{lem:aux_lemma}.    
\end{remark}
\begin{figure}[htp]
\centering
\Gr{\def\x{5*\X}
  \def\y{-7*\Y}
  \foreach \i in {1,...,6} {
    \draw (\X,\y*\i+2.7*\Y) -- (\x*7-\X,\y*\i+2.7*\Y);
    \draw (\x*\i-\X,-0.5*\Y) -- (\x*\i-\X,\y*7+2.9*\Y);
    \place{T\i}{\X cm+\x*\i}{-2*\Y}
    \place{S\i}{2*\X}{\Y cm+\i*\y}
    \foreach \j in {\i,...,6} {
      \Place{\Cl{\j}{\i}}{\x*\i}{\y*\j}
      \pgfmathparse{\x*\i}     \let\xx=\pgfmathresult
      \pgfmathparse{\y*\j-2*\Y}\let\yy=\pgfmathresult
      \place{C\j\i}{\xx}{\yy}
    }
  }}
\caption{\label{fig:tn_contraction}Tensor contraction diagrams for computing $\tr_{23}\left[P^{\sym_3}(I^{\otimes 2}\otimes \ketbra{i}{j})P^{\sym_3}\right]$. The line  represents identity operator,  red box  represents the operator $\ketbra{i}{j}$.}
\end{figure}
\section{Dual SDP for calculating $\eta_k(\rho_{AB})$}\label{app:dual_sdp}
\begin{equation}
    \begin{aligned}
        2^{\eta_k(\rho_{AB})} =\min &\;\; \tr[P_{AB^k}]+\tr[Q_{AB^k}]\\
             {\rm s.t.} &\,\tr_{\backslash {AB_j}}[P_{AB^k}-Q_{AB^k}]=\rho_{AB_j},\, j\in\{1,\cdots,k\}\\
            & \;\; P_{AB^k}\geq 0, Q_{AB^k}\geq 0.
    \end{aligned}
\end{equation}

Without loss of generality, we focus on the case of $k=2$, and derive its dual SDP. Specifically, the Lagrange function is
\begin{align}
    \mathcal{L}(X_{AB_1}, Y_{AB_2}, P_{AB_1B_2}, Q_{AB_1B_2})&:= \tr[P_{AB_1B_2}]+\tr[Q_{AB_1B_2}]\\
    &\quad +\langle X_{AB_1},\,\rho_{AB_1}-\tr_{AB_2}[P_{AB_1B_2}-Q_{AB_1B_2}]\rangle\\
    &\quad +\langle Y_{AB_2},\,\rho_{AB_2}-\tr_{AB_1}[P_{AB_1B_2}-Q_{AB_1B_2}]\rangle\\
    &=\tr[X_{AB_1}\rho_{AB_1}]+\tr[Y_{AB_2}\rho_{AB_2}]\\
    &\quad+ \langle -X_{AB_1}\otimes I_{B_2}, \, P_{AB_1B_2} -Q_{AB_1B_2}\rangle\\
    &\quad+ \langle -Y_{AB_2}\otimes I_{B_1}, \, P_{AB_1B_2} -Q_{AB_1B_2}\rangle\\
    &\quad+\tr[P_{AB_1B_2}]+\tr[Q_{AB_1B_2}]
\end{align}
where $X_{AB_1}$ and $Y_{AB_2}$ are Lagrange multipliers. Then, the Lagrange dual function can be written as
\begin{align}
     \mathcal{G}(X_{AB_1}, Y_{AB_2}):=\inf_{P_{AB_1B_2}\geq0,Q_{AB_1B_2}\geq 0} \mathcal{L}(X_{AB_1}, Y_{AB_2}, P_{AB_1B_2}, Q_{AB_1B_2}).
\end{align}
Since $P_{AB_1B_2}\geq0$ and $Q_{AB_1B_2}\geq0$, it must hold that
\begin{align}
    &I_{AB_1B_2} - X_{AB_1}\otimes I_{B_2} - Y_{AB_2}\otimes I_{B_1} \geq 0,\\
     &I_{AB_1B_2} + X_{AB_1}\otimes I_{B_2} + Y_{AB_2}\otimes I_{B_1} \geq 0,
\end{align}
otherwise, the inner norm is unbounded. Thus, we obtain the following dual SDP
\begin{equation}\label{app:dual_without_rho}
\begin{aligned}
\max &\;\;\tr[X_{AB_1}\rho_{AB_1}]+\tr[Y_{AB_2}\rho_{AB_2}]\\
 {\rm s.t.} 
        & \;\;I_{AB_1B_2} - X_{AB_1}\otimes I_{B_2} - Y_{AB_2}\otimes I_{B_1} \geq 0,\\
        & \;\;I_{AB_1B_2} + X_{AB_1}\otimes I_{B_2} + Y_{AB_2}\otimes I_{B_1} \geq 0.
\end{aligned}
\end{equation}
It is worth noting that the strong duality is held by Slater's condition. Similarly, it is straightforward to generalize it to the case of $k$-extendibility.

\end{document}

%% file: pretex.tex
\usepackage[dvipsnames]{xcolor}
\usepackage{framed}
\definecolor{shadecolor}{rgb}{0.9,0.9,0.9}

\usepackage{mathtools}
\usepackage{amsmath}
\usepackage[shortlabels]{enumitem}

\usepackage{graphicx,epic,eepic,epsfig,amsmath,latexsym,amssymb,verbatim,color}
 
\usepackage{amsfonts}       
\usepackage{nicefrac}       

\usepackage{amsmath}
\usepackage{bbm}

\usepackage{float}
\usepackage{tikz}
\usetikzlibrary{chains}
\usetikzlibrary{fit}
\usepackage{pgflibraryarrows}		
\usepackage{pgflibrarysnakes}		

\usepackage{epsfig}
\usetikzlibrary{shapes.symbols,patterns} 
\usepackage{pgfplots}

\usepackage[strict]{changepage}
\usepackage{hyperref}
\hypersetup{colorlinks=true,citecolor=blue,linkcolor=blue,filecolor=blue,urlcolor=blue,breaklinks=true}

\usepackage[marginal]{footmisc}
\usepackage{url}
\usepackage{theorem}

\newtheorem{definition}{Definition}
\newtheorem{proposition}{Proposition}
\newtheorem{lemma}[proposition]{Lemma}

\newtheorem{theorem}[proposition]{Theorem}

\def\squareforqed{\hbox{\rlap{$\sqcap$}$\sqcup$}}
\def\qed{\ifmmode\squareforqed\else{\unskip\nobreak\hfil
\penalty50\hskip1em\null\nobreak\hfil\squareforqed
\parfillskip=0pt\finalhyphendemerits=0\endgraf}\fi}
\def\endenv{\ifmmode\;\else{\unskip\nobreak\hfil
\penalty50\hskip1em\null\nobreak\hfil\;
\parfillskip=0pt\finalhyphendemerits=0\endgraf}\fi}
\newenvironment{proof}{\noindent \textbf{{Proof~} }}{\hfill $\blacksquare$}

\newcounter{remark}
\newenvironment{remark}[1][]{\refstepcounter{remark}\par\medskip\noindent%
\textbf{Remark~\theremark #1} }{\medskip}

\newcommand{\dketbra}[2]{\vert #1 \rangle \hspace{-.8mm} \rangle \hspace{-.4mm} \langle\hspace{-.8mm}\langle #2 \vert}

\newcommand{\dket}[1]{\vert #1 \rangle \hspace{-.8mm} \rangle}

\newcounter{example}

\mathchardef\ordinarycolon\mathcode`\:
\mathcode`\:=\string"8000
\def\vcentcolon{\mathrel{\mathop\ordinarycolon}}
\begingroup \catcode`\:=\active
  \lowercase{\endgroup
  \let :\vcentcolon
  }

\usepackage{cleveref}
\usepackage{graphicx}
\usepackage{xcolor}

\RequirePackage[framemethod=default]{mdframed}
\newmdenv[skipabove=7pt,
skipbelow=7pt,
backgroundcolor=darkblue!15,
innerleftmargin=5pt,
innerrightmargin=5pt,
innertopmargin=5pt,
leftmargin=0cm,
rightmargin=0cm,
innerbottommargin=5pt,
linewidth=1pt]{tBox}

\newmdenv[skipabove=7pt,
skipbelow=7pt,
backgroundcolor=red!15,
innerleftmargin=5pt,
innerrightmargin=5pt,
innertopmargin=5pt,
leftmargin=0cm,
rightmargin=0cm,
innerbottommargin=5pt,
linewidth=1pt]{rBox}

\newmdenv[skipabove=7pt,
skipbelow=7pt,
backgroundcolor=blue2!25,
innerleftmargin=5pt,
innerrightmargin=5pt,
innertopmargin=5pt,
leftmargin=0cm,
rightmargin=0cm,
innerbottommargin=5pt,
linewidth=1pt]{dBox}
\newmdenv[skipabove=7pt,
skipbelow=7pt,
backgroundcolor=darkkblue!15,
innerleftmargin=5pt,
innerrightmargin=5pt,
innertopmargin=5pt,
leftmargin=0cm,
rightmargin=0cm,
innerbottommargin=5pt,
linewidth=1pt]{sBox}
\definecolor{darkblue}{RGB}{0,76,156}
\definecolor{darkkblue}{RGB}{0,0,153}
\definecolor{blue2}{RGB}{102,178,255}
\definecolor{darkred}{RGB}{195,0,0}

\newcommand{\nc}{\newcommand}
\nc{\rnc}{\renewcommand}
\nc{\lbar}[1]{\overline{#1}}
\nc{\bra}[1]{\langle#1|}
\nc{\ket}[1]{|#1\rangle}
\nc{\ketbra}[2]{|#1\rangle\!\langle#2|}
\nc{\braket}[2]{\langle#1|#2\rangle}

\nc{\proj}[1]{| #1\rangle\!\langle #1 |}
\nc{\avg}[1]{\langle#1\rangle}
\nc{\smfrac}[2]{\mbox{$\frac{#1}{#2}$}}
\nc{\tr}{\operatorname{Tr}}
\nc{\ox}{\otimes}
\nc{\dg}{\dagger}
\nc{\dn}{\downarrow}
\nc{\cA}{{\cal A}}
\nc{\cB}{{\cal B}}
\nc{\cC}{{\cal C}}
\nc{\cD}{{\cal D}}
\nc{\cE}{{\cal E}}
\nc{\cF}{{\cal F}}
\nc{\cG}{{\cal G}}
\nc{\cH}{{\cal H}}
\nc{\cI}{{\cal I}}
\nc{\cJ}{{\cal J}}
\nc{\cK}{{\cal K}}
\nc{\cL}{{\cal L}}
\nc{\cM}{{\cal M}}
\nc{\cN}{{\cal N}}
\nc{\cO}{{\cal O}}
\nc{\cP}{{\cal P}}
\nc{\cQ}{{\cal Q}}
\nc{\cR}{{\cal R}}
\nc{\cS}{{\cal S}}
\nc{\cT}{{\cal T}}
\nc{\cU}{{\cal U}}
\nc{\cV}{{\cal V}}
\nc{\cX}{{\cal X}}
\nc{\cY}{{\cal Y}}
\nc{\cZ}{{\cal Z}}
\nc{\cW}{{\cal W}}
\nc{\csupp}{{\operatorname{csupp}}}
\nc{\qsupp}{{\operatorname{qsupp}}}
\nc{\var}{{\operatorname{var}}}
\nc{\rar}{\rightarrow}
\nc{\lrar}{\longrightarrow}
\nc{\polylog}{{\operatorname{polylog}}}
\nc{\wt}{{\operatorname{wt}}}
\nc{\av}[1]{{\left\langle {#1} \right\rangle}}
\nc{\supp}{{\operatorname{supp}}}

\nc{\argmin}{{\operatorname{argmin}}}

\def\i{\mathbf{i}}

\def\x{\xi}

\nc{\RR}{{{\mathbb R}}}
\nc{\CC}{{{\mathbb C}}}
\nc{\FF}{{{\mathbb F}}}
\nc{\NN}{{{\mathbb N}}}
\nc{\ZZ}{{{\mathbb Z}}}
\nc{\PP}{{{\mathbb P}}}
\nc{\QQ}{{{\mathbb Q}}}
\nc{\UU}{{{\mathbb U}}}
\nc{\EE}{{{\mathbb E}}}
\nc{\id}{{\operatorname{id}}}

\nc{\CHSH}{{\operatorname{CHSH}}}

\nc{\be}{\begin{equation}}
\nc{\ee}{{\end{equation}}}
\nc{\bea}{\begin{eqnarray}}
\nc{\eea}{\end{eqnarray}}
\nc{\<}{\langle}
\rnc{\>}{\rangle}
\nc{\rU}{\mbox{U}}

\nc{\ob}[1]{#1}

\nc{\SEP}{{\text{\rm SEP}}}
\nc{\NS}{{\text{\rm NS}}}
\nc{\LOCC}{{\text{\rm LOCC}}}
\nc{\PPT}{{\text{\rm PPT}}}
\nc{\EXT}{{\text{\rm EXT}}}
\nc{\Sym}{{\operatorname{Sym}}}

\nc{\ERLO}{{E_{\text{r,LO}}}}
\nc{\ERLOCC}{{E_{\text{r,LOCC}}}}
\nc{\ERPPT}{{E_{\text{r,PPT}}}}
\nc{\ERLOCCinfty}{{E^{\infty}_{\text{r,LOCC}}}}
\nc{\Aram}{{\operatorname{\sf A}}}

\usepackage{tikz}
\usepackage{hyperref}
\hypersetup{colorlinks=true,citecolor=blue,linkcolor=blue,filecolor=blue,urlcolor=blue,breaklinks=true}

\makeatletter
\def\grd@save@target#1{%
  \def\grd@target{#1}}
\def\grd@save@start#1{%
  \def\grd@start{#1}}
\tikzset{
  grid with coordinates/.style={
    to path={%
      \pgfextra{%
        \edef\grd@@target{(\tikztotarget)}%
        \tikz@scan@one@point\grd@save@target\grd@@target\relax
        \edef\grd@@start{(\tikztostart)}%
        \tikz@scan@one@point\grd@save@start\grd@@start\relax
        \draw[minor help lines,magenta] (\tikztostart) grid (\tikztotarget);
        \draw[major help lines] (\tikztostart) grid (\tikztotarget);
        \grd@start
        \pgfmathsetmacro{\grd@xa}{\the\pgf@x/1cm}
        \pgfmathsetmacro{\grd@ya}{\the\pgf@y/1cm}
        \grd@target
        \pgfmathsetmacro{\grd@xb}{\the\pgf@x/1cm}
        \pgfmathsetmacro{\grd@yb}{\the\pgf@y/1cm}
        \pgfmathsetmacro{\grd@xc}{\grd@xa + \pgfkeysvalueof{/tikz/grid with coordinates/major step}}
        \pgfmathsetmacro{\grd@yc}{\grd@ya + \pgfkeysvalueof{/tikz/grid with coordinates/major step}}
        \foreach \x in {\grd@xa,\grd@xc,...,\grd@xb}
        \node[anchor=north] at (\x,\grd@ya) {\pgfmathprintnumber{\x}};
        \foreach \y in {\grd@ya,\grd@yc,...,\grd@yb}
        \node[anchor=east] at (\grd@xa,\y) {\pgfmathprintnumber{\y}};
      }
    }
  },
  minor help lines/.style={
    help lines,
    step=\pgfkeysvalueof{/tikz/grid with coordinates/minor step}
  },
  major help lines/.style={
    help lines,
    line width=\pgfkeysvalueof{/tikz/grid with coordinates/major line width},
    step=\pgfkeysvalueof{/tikz/grid with coordinates/major step}
  },
  grid with coordinates/.cd,
  minor step/.initial=.2,
  major step/.initial=1,
  major line width/.initial=2pt,
}
\makeatother

\usepackage{thmtools}
\usepackage{thm-restate}
\usepackage{etoolbox}
\makeatletter
\def\problem@s{}
\newcounter{problems@cnt}

\newcommand{\allproblems}{\problem@s}
\makeatother

\newcommand{\Gr}[2][-0.1cm]{
\begin{tikzpicture}[semithick, baseline = #1,
  box/.style = {draw, fill = white, inner sep = 0, minimum height = 9, minimum width = 9},
  tri/.style = {box, inner xsep = 1.5, isosceles triangle}]
  \newcommand{\X}{0.50}
  \newcommand{\Y}{0.45}
  #2
\end{tikzpicture}}

\newcommand{\Place}[3]{
\begin{scope}[xshift = #2cm, yshift = #3cm]
  #1
\end{scope}}
\newcommand{\place}[3]{\Place{\csuse{#1}}{#2}{#3}}

\newcommand{\round}{.. controls +(0.3,0) and +(-0.3,0) ..}

\csdef{S1}{
  \draw (0, \Y)   --   (\X, \Y);
  \draw (0,  0)   --   (\X,  0);
  \draw (0,-\Y)   --   (\X,-\Y);
}
\csdef{S2}{
  \draw (0, \Y) \round (\X,  0);
  \draw (0,  0) \round (\X,-\Y);
  \draw (0,-\Y) \round (\X, \Y);
}
\csdef{S3}{
  \draw (0, \Y) \round (\X,-\Y);
  \draw (0,  0) \round (\X, \Y);
  \draw (0,-\Y) \round (\X,  0);
}
\csdef{S4}{
  \draw (0, \Y)   --   (\X, \Y);
  \draw (0,  0) \round (\X,-\Y);
  \draw (0,-\Y) \round (\X,  0);
}
\csdef{S5}{
  \draw (0, \Y) \round (\X,  0);
  \draw (0,  0) \round (\X, \Y);
  \draw (0,-\Y)   --   (\X,-\Y);
}
\csdef{S6}{
  \draw (0, \Y) \round (\X,-\Y);
  \draw (0,  0)   --   (\X,  0);
  \draw (0,-\Y) \round (\X, \Y);
}

\csdef{T1}{\csuse{S1}}
\csdef{T2}{\csuse{S3}} 
\csdef{T3}{\csuse{S2}} 
\csdef{T4}{\csuse{S4}}
\csdef{T5}{\csuse{S5}}
\csdef{T6}{\csuse{S6}}

\csdef{Bx}#1#2#3{
  \place{S1}{1*\X}{0}
  \node[box] at (1.5*\X,-\Y) {\scriptsize };
}
\csdef{Bxs}{\Bx{1}{2}{3}}

\csdef{Tri}#1#2#3{
  \place{S1}{1*\X}{0}
  \node[tri] at (1.5*\X, \Y) {\scriptsize #1};
  \node[tri] at (1.5*\X,  0) {\scriptsize #2};
  \node[tri] at (1.5*\X,-\Y) {\scriptsize #3};
}

\newcommand{\db}{0.12} 
\newcommand{\loopr}{0.15} 
\newcommand{\loopR}{0.35} 

\newcommand{\trloop}[4]{
  \coordinate (I3) at (#1*\X,#3*\Y);  \coordinate (L3) at ($(I3)+(0,-0.4*\Y-\db)$);
  \coordinate (O3) at (#2*\X,#3*\Y);  \coordinate (R3) at ($(O3)+(0,-0.4*\Y-\db)$);
  \draw (I3) ..controls +(-\loopr,0) and +(-\loopr,0) .. (L3)
     -- (R3) ..controls +( \loopr,0) and +( \loopr,0) .. (O3)#4; 
}

\newcommand{\Trloop}[3]{
  \coordinate (I2) at (#1*\X,#3*\Y);  \coordinate (L2) at ($(I2)+(0,-1.4*\Y-2*\db)$);
  \coordinate (O2) at (#2*\X,#3*\Y);  \coordinate (R2) at ($(O2)+(0,-1.4*\Y-2*\db)$);
  \draw (I2) ..controls +(-\loopR,0) and +(-\loopR,0) .. (L2)
     -- (R2) ..controls +( \loopR,0) and +( \loopR,0) .. (O2);
}

\newcommand{\Cl}[2]{
  \place{S#1}{0*\X}{1*\Y}
  \place{Bxs}{0*\X}{1*\Y}
  \place{T#2}{2*\X}{1*\Y}
  \draw (-0.4*\X,2*\Y) -- (     0,2*\Y);
  \draw (   3*\X,2*\Y) -- (3.4*\X,2*\Y);
  \trloop{0}{3}{0}{}
  \Trloop{0}{3}{1}
}

\newcommand{\IOO}[3]{
  \draw (-0.2*\X,0) -- (3.2*\X,0);
  \node[circle] at (1.5*\X,  0) {\scriptsize  };
  \node[circle] at (0.5*\X,-\Y) {\scriptsize };
  \node at (1.5*\X,-\Y) {\scriptsize $d\delta_{ij}$};
}
\csdef{C11}{\IOO{1}{2}{3}}
\csdef{C22}{\IOO{2}{1}{3}}
\csdef{C44}{\IOO{1}{2}{3}}
\csdef{C55}{\IOO{1}{2}{3}}

\newcommand{\IOOA}[3]{
  \draw (-0.2*\X,0) -- (3.2*\X,0);
  \node[box, fill=red!30] at (1.5*\X,  0) {\scriptsize  };
  \node[circle] at (0.5*\X,-\Y) {\scriptsize };
  \node[circle] at (1.5*\X,-\Y) {\scriptsize $d^2$ };

}
\csdef{C33}{\IOOA{1}{2}{3}}
\csdef{C66}{\IOOA{1}{2}{3}}

\newcommand{\IIO}[3]{
  \draw (-0.2*\X,0) -- (3.2*\X,0);
  \node[box, fill=red!30] at (  1.5*\X,  0) {\scriptsize };
 \node[circle] at (1.5*\X,-\Y) {\scriptsize $d$};
  \node[circle] at (2.5*\X,-\Y) {\scriptsize };
}

\csdef{C61}{\IIO{2}{1}{3}}
\csdef{C43}{\IIO{2}{1}{3}}
\csdef{C53}{\IIO{2}{1}{3}}
\csdef{C62}{\IIO{2}{1}{3}}

\newcommand{\IIOA}[3]{
  \draw (-0.2*\X,0) -- (3.2*\X,0);
  \node[circle] at (  1.5*\X,  0) {\scriptsize };
 \node[circle] at (0.5*\X,-\Y) {\scriptsize };
  \node[circle] at (1.5*\X,-\Y) {\scriptsize $\delta_{ij}$};
}
\csdef{C51}{\IIOA{1}{2}{3}}
\csdef{C42}{\IIOA{1}{2}{3}}

\newcommand{\OOO}[3]{
  \draw (-0.2*\X,0) -- (3.2*\X,0);
  \node[circle] at (1.5*\X,  0) {\scriptsize };
  \node[circle] at (  1*\X,-\Y) {\scriptsize };
  \node[circle] at (  1.5*\X,-\Y) {\scriptsize $\delta_{ij}$};
}
\csdef{C41}{\OOO{1}{2}{3}}
\csdef{C52}{\OOO{1}{2}{3}}

\newcommand{\OOOA}[3]{
  \draw (-0.2*\X,0) -- (3.2*\X,0);
  \node[box, fill=red!30] at (1.5*\X,  0) {\scriptsize };
  \node[circle] at (  1*\X,-\Y) {\scriptsize };
  \node[circle] at (  1.5*\X,-\Y) {\scriptsize $d$};
}
\csdef{C63}{\OOOA{1}{2}{3}}

\newcommand{\III}[3]{
  \draw (-0.2*\X,-0.5*\Y) -- (3.2*\X,-0.5*\Y);
  \node[box, fill=red!30] at (1.5*\X,-0.5*\Y) {\scriptsize };
}
\csdef{C21}{\III{2}{3}{1}}
\csdef{C31}{\III{3}{2}{1}}
\csdef{C32}{\III{3}{1}{2}}
\csdef{C54}{\III{2}{3}{1}}
\csdef{C64}{\III{3}{2}{1}}
\csdef{C65}{\III{3}{1}{2}}